\documentclass[%
 reprint,
superscriptaddress,
 amsmath,amssymb,
 aps,
prb,
]{revtex4-2}

\usepackage{placeins}
\usepackage{afterpage}
\usepackage{graphicx}
\usepackage{dcolumn}
\usepackage{bm}
\usepackage[colorlinks=true, allcolors=blue]{hyperref}

\usepackage{xcolor}

\usepackage{physics}
\usepackage{acro}
\usepackage{caption}
\usepackage{subcaption}
\usepackage{lipsum}
\usepackage[compat=1.1.0]{tikz-feynman} 
\usepackage{feynmp-auto}
\DeclareAcronym{BCS}{
    short=BCS,
    long=Bardeen–Cooper–Schrieffer
}

\DeclareAcronym{C2DB}{
    short=C2DB,
    long=computational 2D material database
}

\DeclareAcronym{cRPA}{
    short=cRPA,
    long=constrained Random Phase Approximation,
}

\DeclareAcronym{DFT}{
    short=DFT,
    long=density functional theory
}

\DeclareAcronym{hBN}{
    short=hBN,
    long=hexagonal Boron-Nitride
}

\DeclareAcronym{RPA}{
    short=RPA,
    long=random phase approximation
}

\DeclareAcronym{TMD}{
    short=TMD,
    long=transition metal dichalcogenide
}

\DeclareAcronym{TMO}{
    short=TMO,
    long=transition metal oxide
}

\DeclareAcronym{vdWH}{
    short=vdWH,
    long=van der Waals heterostructure,
}

\DeclareAcronym{QEH}{
    short=QEH,
    long=quantum-electrostatic heterostructure
}

\DeclareAcronym{QP}{
    short=QP,
    long=quasi particle
}

\begin{document}

\preprint{APS/123-QED}

\title{Exciton Superfluidity in 2D Heterostructures from First Principles: The importance of material specific screening}

\author{Rune H\o jlund}
\affiliation{These authors contributed equally to this work.}
\affiliation{
Center for Atomic-scale Materials Design,
Department of Physics,
Technical University of Denmark,
DK-2800 Kgs. Lyngby, Denmark
}
\author{Emil Grovn}
\affiliation{These authors contributed equally to this work.}
\affiliation{
Center for Atomic-scale Materials Design,
Department of Physics,
Technical University of Denmark,
DK-2800 Kgs. Lyngby, Denmark
}
\author{Sahar Pakdel}
\affiliation{
Center for Atomic-scale Materials Design,
Department of Physics,
Technical University of Denmark,
DK-2800 Kgs. Lyngby, Denmark
}
\author{Kristian S. Thygesen}
\affiliation{
Center for Atomic-scale Materials Design,
Department of Physics,
Technical University of Denmark,
DK-2800 Kgs. Lyngby, Denmark
}
\author{Fredrik Nilsson}
\affiliation{
Center for Atomic-scale Materials Design,
Department of Physics,
Technical University of Denmark,
DK-2800 Kgs. Lyngby, Denmark
}

\date{\today}

\begin{abstract}
Recent theoretical and experimental studies suggest that van der Waals heterostructures with n- and p-doped bilayers of transition metal dichalcogenides are promising facilitators of exciton superfluidity. Exciton superfluidity in such bilayer systems is often modelled by solving a mean-field gap equation defined for only the conduction and valence band of the electron and hole material respectively. A key quantity entering the gap equation is the effective Coulomb potential acting as the bare interaction in the subspace of the model. Since the model only includes a few bands around the Fermi energy the effective model interaction is partially screened. Although the screening is a material dependent quantity it has in previous works been accounted for in an \emph{ad hoc} manner, by assuming a static dielectric constant of 2 for a wide range of different materials. In this work we show that the effective model interaction can be derived from first principles using open source code frameworks. Using this novel \emph{ab initio} downfolding procedure we show that the material dependent screening is essential to predict both magnitude and trends of exciton binding energies and superfluid properties. 
Furthermore, we suggest new material platforms of both transition metal oxides and dichalcogenides with superior properties compared to the standard devices with two transition metal dichalcogenide layers.
\end{abstract}

\maketitle

\section{Introduction}
Excitons are quasi-particle (QP) excitations of electron-hole pairs bound by their Coulomb attraction \cite{Ueta1986}. Since these bound electron-hole pairs are approximately bosonic quasiparticles, they may, under special circumstances, form a highly correlated condensate state with exotic properties such as superfluidity \cite{Lozovik2009}.
In this work we consider spatially indirect excitons in \acp{vdWH}. \acp{vdWH} is a class of materials consisting of stacked monolayers of two-dimensional (2D) atomic crystals \cite{Geim2013}. The individual layers in the \ac{vdWH} may act as quasi-2D quantum wells and by doping, the structures can be engineered to carry electrons in one layer and holes in a separate layer. Since the electrons and holes are spatially separated, the exciton condensation can give rise to counter-flowing superconductivity \cite{Lozovik1976}, though the concept of superconductivity here is delicate, since no net charge transport is achieved \cite{Fil2018}. From a theoretical perspective, these bilayer architectures are interesting as potential tunable model systems of the BCS-BEC (Bardeen–Cooper–Schrieffer––Bose–Einstein-condensate) cross-over. 
The iron pnictides are believed to be located at the verge of the BCS-BEC crossover \cite{kasahara2014}, and an increased understanding of this regime could yield important new clues to the mechanism behind the unconventional superconductivity in these compounds.
Along with the rapid development in synthesis of 2D materials, different elements and structures have been considered as facilitators of the phenomenon: from GaAs heterostructures, \cite{Zhu1995, Croxall2008, Seamons2009}, to double mono- and bilayer graphene, \cite{Min2008, Lozovik2009, Gorbachev2012, Lozovik2012, Perali2013, Burg2018, Nilsson2021}, and most recently bilayers of \acp{TMD} \cite{bi2021, ma2021, Conti2020, Conti2020b, Wang2019, Conti2019, Conti2020, berman2016, wu2015, fogler2014}. 

Experimentally, spectroscopic indications of exciton condensation has been seen in both bulk materials, such as 1T-TiSe$_2$ \cite{Cercellier2007, kogar2017} and Ta$_2$NiSe$_5$ \cite{Seki2014} and in InAs/GaSb bilayers \cite{du2017}. Two-dimensional heterostructures consisting of two semiconducting sheets separated by a dielectric barrier has shown to be a particularly convenient, tunable platform to study exciton condensation and exciton superfluidity \cite{xie2018, Burg2018, Wang2019, Zeng2020, ma2021}. The recent experimental indications of interlayer exciton condensation at strikingly high temperatures (up to ~100 K) in \ac{TMD} bilayer structures \cite{Wang2019, ma2021} is of particular interest to this work. 

Exciton superfluidity in bilayer systems is often modelled by solving a mean-field gap equation defined for only the conduction and valence band of the electron and hole material respectively. 
From a theoretical point of view the screening within this low-energy subspace has been identified as a key quantity that will reduce the electron-hole interaction and thus severely reduce the superfluid transition temperature \cite{Nilsson2021, Conti2019, berman2017}. 
The quenching of the screening in the condensate state is manifested by the partial cancellation between the real and anomalous screening channels. Over the years accurate methods to account for these competing screening channels have been developed \cite{Nilsson2021, Perfetto2020, Lozovik2012, lozovik1997} and used to describe and predict the superfluid properties of many different systems \cite{Nilsson2023, pascucci2022, Nilsson2021, gupta2020,  Perfetto2020, Conti2020, Conti2020b, Conti2019, Conti2017, Debnath2017, Zarenia2014, Perali2013, Lozovik2012, lozovik1996}. 

However, real materials have many screening channels that are not explicitly accounted for when solving such low-energy models. From a model perspective all such additional screening channels are accounted for by the reduction (screening) of the interaction parameters of the model Hamiltonian. This screening is in fact a material dependent quantity but has in previous works \cite{pascucci2022, Nilsson2021, gupta2020,  Conti2020, Conti2020b, Conti2019, Conti2017, Debnath2017, wu2015, Zarenia2014, Perali2013, Lozovik2012, lozovik1996}
been accounted for in an \emph{ad hoc} manner by assuming a single static dielectric constant, often chosen as $\epsilon^i_M = 2$ for a wide range of different materials \cite{Conti2019, Conti2020, Nilsson2021, Lozovik2012}.
Recently a novel \emph{ab initio} method to compute the effective model parameters was suggested and applied to investigate exciton condensation in Janus bilayer materials \cite{Nilsson2023}. In this work we give a detailed derivation of this method and show that the effective model interaction can be derived from first principles using open source frameworks. Through a systematic investigation of TMD and \ac{TMO} bilayers we show that the material dependent screening is essential to predict both magnitude and trends of exciton binding energies and superfluid properties. Using this method we suggest new material platforms with superior properties compared to the standard devices.
The objective of this study is thus two-fold:\\
\begin{figure*}[ht!]
    \centering
    \vspace{1em}
    \includegraphics[width=0.62\linewidth]{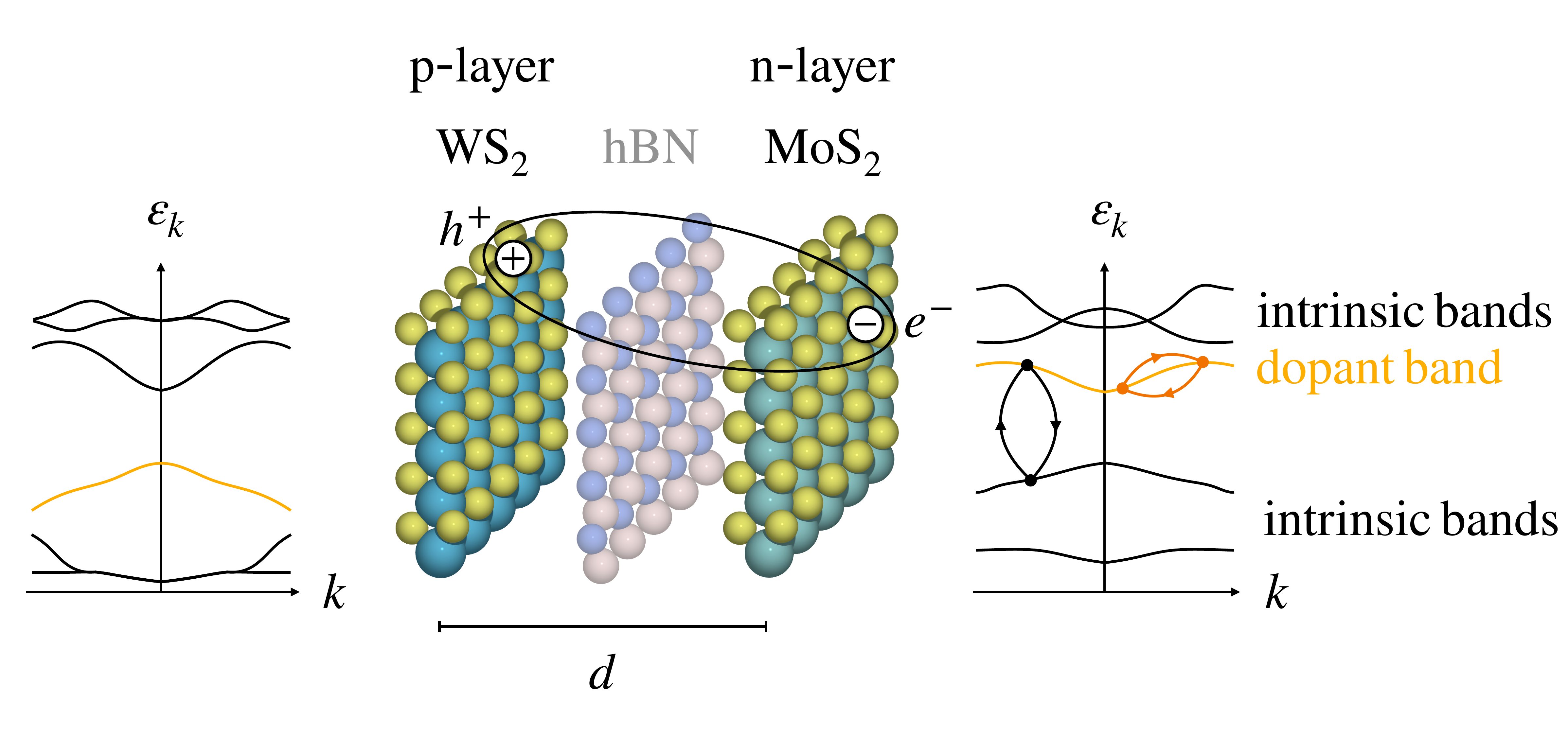}
    \caption{Example sketch of a TMD bilayer structure with an n-doped $\mathrm{MoS_2}$ layer and a p-doped $\mathrm{WS_2}$-layer. An indirect exciton (electron-hole pair) is shown schematically. We have indicated the partitioning of the band structures of each of the layers into the \textbf{dopant} (valence/conduction) band and remaining \textbf{intrinsic} bands as explained in Section \ref{sec:cRPA}.}
    \label{fig:bilayer}
\end{figure*}
i) In the quest for better material candidates, this study offers a computational high-throughput investigation of novel bilayer structures. As 2D layer materials we consider both the previously explored \acp{TMD}, but we also expand the search to include new \acp{TMD} as well as \acp{TMO}.
\\
ii) We give a detailed derivation of the \emph{ab initio} method suggested in Ref. \cite{Nilsson2023} and make a systematic study for a large number of TMD and TMO platforms. 

The article is organized as follows: 
In Section \ref{sec:methods} we introduce the specific material platforms (vdW Heterostructures) that we consider in this work and give a detailed account for the theoretical methods we use to model them. We start by providing an overview of the workflow for the material screening in Section \ref{sec:workflow}-\ref{sec:selecting-materials}. Then we provide an intuitive derivation of the gap equation used to model the superfluid state in Section \ref{sec:gap} and the novel downfolding approach used to compute the effective parameters of the gap equation in Section \ref{sec:screening}- \ref{sec:Mott-Wannier}. Finally we present the results in Section \ref{sec:results} and in Section \ref{sec:conclusions} we summarize the most important conclusions.

\section{Methods \label{sec:methods}}

\subsection{Material platform}
An example of the bilayer structures we consider is shown in Fig.~\ref{fig:bilayer}. This is a \ac{vdWH}, where we have an n-doped and a p-doped \ac{TMD} monolayer, spatially separated by a \ac{hBN} layer. The bilayer combination in Fig.~\ref{fig:bilayer} is just one example of the 336 possible combinations for 27 different \acp{TMD} and \acp{TMO} which we consider in this work. Besides the structure itself, Fig.~\ref{fig:bilayer} also indicates (schematically) part of the bandstructure close to the partially occupied conduction and valence band (denoted "dopant band") for the n- and p-doped layer respectively. The construction allows electrons (mainly from the conduction band) of the n-layer to interact through the (screened) Coulomb interaction with holes (mainly from the valence band) of the p-layer, thereby forming spatially indirect excitons. Similar to the experimental setup in Ref. \cite{Wang2019} the individual electron and hole concentrations can be tuned using external gates.

\subsection{Methodology/Workflow \label{sec:workflow}}
\begin{figure*}[htb]
    \centering
    \includegraphics[width=1.0\textwidth]{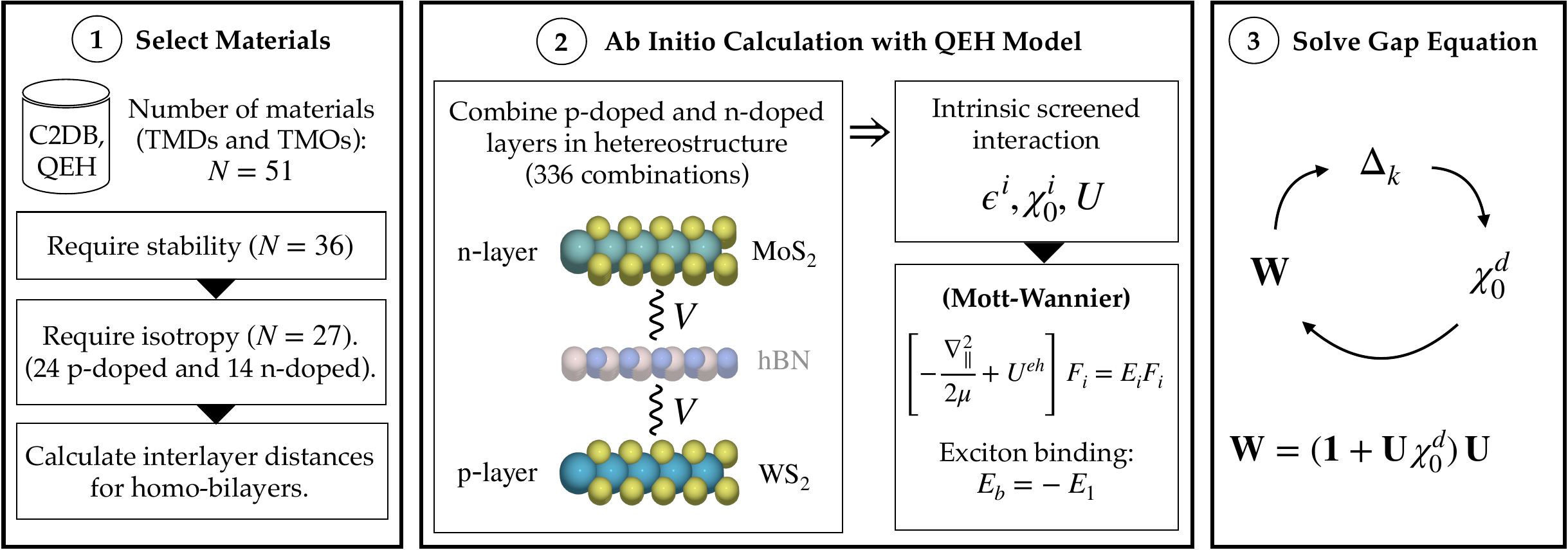}
    \caption{Overview of the employed workflow. First we screen the QEH 2D material database for materials that fulfill the stability and bandstructure criteria (according to the data extracted from the Computational 2D Material Database (C2DB)). In the second step the monolayers are combined to heterostructures and the intra and interlayer Coulomb interactions are computed. These are used to calculate the exciton binding energies using a Wannier Mott model and as \emph{ab initio} inputs to the gap equation in step 3.}
    \label{fig:workflow}
\end{figure*}

In Fig.~\ref{fig:workflow} we show an overview of the workflow used in this computational study. In the first step we selected stable 2D monolayers to use as the individual layers of the \acp{vdWH} (see Section \ref{sec:selecting-materials}). In the second step, the monolayers are combined in \acp{vdWH} and we calculated the dielectric function, polarization and screened interaction of the undoped \acp{vdWH} using \emph{ab initio} methods (see Section \ref{sec:QEH}). In the final step we used the \emph{ab initio} results as input for solving a mean-field gap equation defined only for the conduction and valence band of $n$- and $p$-doped materials respectively (see Section \ref{sec:gap}).

\subsection{Selecting Materials \label{sec:selecting-materials}}
Our computational study is based on the \ac{QEH} 2D Material Database, Ref. \cite{Andersen2015, QEH_DB}, which contains precalculated dielectric properties (response functions and induced densities) for 51 \acp{TMD} and \acp{TMO}. We imposed a number of requirements for inclusion in the study:
\begin{enumerate}
    \item Thermodynamic stability: We only considered dynamically and thermodynamically stable materials with a maximum energy of 0.2 eV above the convex Hull, as given in the \ac{C2DB}\cite{Haastrup2018, Gjerding2021}.
    \item Band structure requirements: In order for hBN to function as a barrier we only considered the n-doped materials with the conduction band minimum (cbm) in the hBN bandgap. Correspondingly we required the valence band maximum (vbm) of the p-doped materials to be above the vbm of hBN. These requirements were verified from the GW bandstructures in the \ac{C2DB}. For computational simplicity we only considered isotropic materials, i.e. materials for which the effective masses within 2 directions around the band-extremum are equal within a 2 \% tolerance. For p-doped materials we required isotropy in the valence band. For n-doped materials we required isotropy in the conduction band.
\end{enumerate}
The requirements above left us with 27 unique materials of which 24 were suitable for n-doping and 14 were suitable for p-doping yielding a total of $24 \times 14 = 336$ possible bilayer combinations to investigate.

After the material selection we estimated the "thickness" of each layer in the \ac{vdWH} using \ac{DFT} calculations in a $z$-scan approach \cite{Pakdel2023}.
The $z$-scan approach evaluates the interlayer distance and binding energy of a homobilayer by calculating the total energy of the bilayer while varying the distance between two monolayers for a given lateral stacking configuration. In the present work the total energy is evaluated using the PBE-D3 xc-functional. We consider all possible stacking configurations and determine the interlayer distance from the most stable stacking. For a heterostructure we approximate the interlayer distance between layer A and layer B by the average of the interlayer distance calculated for homobilayers AA and BB, respectively. It should be noted that in our calculations the interlayer distance is the only stacking dependent quantity since we determine the single-particle dispersion from the effective mass of the monolayers and the interaction parameters using the \ac{QEH} model (see Sections IID-H).

\subsection{The gap equation \label{sec:gap}}
Exciton superfluidity in bilayer systems is often modeled within mean field BCS theory derived from a Hamiltonian spanned by a few relevant bands, typically chosen as the conduction and valence band of the electron and hole material respectively \cite{Conti2020, Nilsson2021, Lozovik2009, Lozovik2012}.  As indicated in Fig.~\ref{fig:bilayer} we therefore consider a model of two bands; the conduction band of the n-doped material ($n$=1) and the valence band of the p-doped material ($n$=2). For the materials  with non-negligable spin-orbit coupling (SOC) we consider both spin-orbit split bands in the low-energy model. The momentum dependence of the spin-orbit splitting for the materials in our study is weak. Therefore, the SOC is parameterized by a single number, similar to Refs. \onlinecite{Conti2020, pascucci2022, Conti2020b}. Within a parabolic approximation the bare dispersion of the electron and hole bands are then given by (in atomic units)

\begin{align}
\varepsilon^{(n)}_{\vb{k}\gamma} = \frac{k^2}{2m^{(n)}} + \gamma  \frac{\lambda_{\mathrm{SOC}}}{2}
\end{align}
where $n$ is the layer index, $\lambda_{\mathrm{SOC}}$ is the spin-orbit splitting of the conduction (valence) band in the electron (hole) layers, $\gamma = \pm 1$ labels the two spin-orbit split bands and the effective masses $m^{(n)}$ are taken from the \ac{C2DB}.
To simplify the notation we will in the remainder of this section restrict the discussion to the case where $\lambda_{\mathrm{SOC}}=0$ and drop the index $\gamma$, noting that the generalization to finite SOC is straight-forward. 

The resulting many-body Hamiltonian is:
\begin{subequations}
\label{eq:H-full}
\begin{gather}
H
= H_0 + \frac{1}{2}
\sum_{
\substack{
\sigma, \sigma', \, n, n', \\
\vb{k}, \vb{k'}, \, \vb{q}
}}
U^{(n n')}_{\vb{qkk}'}
c^{(n) \dag}_{\vb{k} + \vb{q} \sigma}
c^{(n') \dag}_{\vb{k}' - \vb{q} \sigma'}
c^{(n')}_{\vb{k}' \sigma'}
c^{(n)}_{\vb{k} \sigma}, \\
\text{where} \qquad
H_0 = \sum_{\sigma, n, \vb{k}} \xi^{(n)}_{\vb{k}} c^{(n) \dag}_{\vb{k} \sigma} c^{(n)}_{\vb{k} \sigma}.
\end{gather}
\end{subequations}
Here $c^{(n) \dag}_{\vb{k} \sigma}$ is the creation operator for an electron with spin $\sigma$ in layer $n$, $\xi^{(n)}_{\vb{k}} = \varepsilon^{(n)}_{\vb{k}} - \mu^{(n)}_{\vb{k}}$ is the corresponding kinetic energy adjusted by the chemical potential $\mu^{(n)}_{\vb{k}}$, and $U^{(nn')}_{\vb{q}}$ is a Coulomb interaction. An important point is that we distinguish between the "bare" Coulomb interaction, $V$, the "effective" interaction of the subspace of our superfluidity model, $U$, and the "fully screened" interaction, $W$. As we shall see, the fully screened interaction, $W$ is what enters our final gap equation. $U$ is a partially screened interaction which is screened by all screening channels not explicitly accounted for in the low-energy model in Eq.~\eqref{eq:H-full}. $U$ can be implicitly defined by enforcing the fully screened interaction calculated from the model Hamiltonian in Eq.~\eqref{eq:H-full} to be equivalent to the fully screened interaction computed for the full many-body Hamiltonian defined in the complete Hilbert space. In principle $U$ is a frequency dependent quantity, but we only consider the static limit of $U$ in this work. We will return to the calculation of $U$ and $W$ in Sections \ref{sec:screening} - \ref{sec:QEH}.

\subsubsection{BCS Mean Field Model}
\noindent To focus on the most important dynamics of exciton superfluidity, we will begin by considering a simpler Hamiltonian containing only the electron-hole interaction between the n-doped layer (layer 1) and p-doped layer (layer 2):
\begin{gather}
\label{eq:H-U-e-h}
H
= H_0 +
\sum_{\vb{k}, \vb{k'}}
U^{(1 2)}_{\vb{k}'-\vb{k}} \,
c^{(1) \dag}_{\vb{k}' \uparrow}
c^{(2)}_{\vb{k}' \downarrow}
c^{(2) \dag}_{\vb{k} \downarrow}
c^{(1)}_{\vb{k} \uparrow},
\end{gather}
Note that $c^{(1) \dag}_{\vb{k}' \uparrow}$ creates an electron with momentum $\vb{k}'$ in layer 1, whereas $c^{(2)}_{\vb{k}' \downarrow}$ creates a hole with momentum $-\vb{k}'$ in layer 2. In this simpler Hamiltonian, we assume that electron-hole pairs form with opposite spin and momentum (singlet phase-coherent coupling). Under this assumption, spin will be dropped from the equations for brevity. In BCS theory it is assumed that electron-hole pairs will form a coherent groundstate and therefore the pair mean-field
$\expval{c^{(1) \dag}_{\vb{k}} c^{(2)}_{\vb{k}} }$ is assumed to be non-zero. By assuming the deviation from the mean-field, $c^{(1) \dag}_{\vb{k}} c^{(2)}_{\vb{k}}  - \expval{c^{(1) \dag}_{\vb{k}} c^{(2)}_{\vb{k}} }$, to be small, we can simplify the Hamiltonian (constant terms are absorbed by the chemical potential in $H_0$):
\begin{gather}
\label{eq:H-BCS}
H
= H_0 
- \sum_{\vb{k}} \qty(
\Delta_{\vb{k}} c^{(1) \dag}_{\vb{k}} c^{(2)}_{\vb{k}}
+ \Delta^*_{\vb{k}} c^{(2) \dag}_{\vb{k}} c^{(1)}_{\vb{k}} ),
\end{gather}
where we have introduced the order parameter:
\begin{gather}
\label{eq:gap-1}
\Delta_{\vb{k}}
= - \sum_{\vb{k}'} U^{(12)}_{\vb{k' - k}}
\expval{ c^{(1)}_{\vb{k'}} c^{(2) \dag}_{\vb{k'}} }.
\end{gather}
The Hamiltonian can be written on matrix form by introducing the Nambu-spinors:
\begin{gather*}
    \psi_{\vb{k}} =
    \begin{pmatrix}
        c^{(1)}_{\vb{k}} \\[0.3em]
        c^{(2)}_{\vb{k}}
    \end{pmatrix}, \,\, \psi^\dag_{\vb{k}} =
    \begin{pmatrix}
        c^{(1) \dag}_{\vb{k}} && \!\!\!\!
        c^{(2) \dag}_{\vb{k}}
    \end{pmatrix}, \,\,
    \qty{\psi_{\vb{k} i}, \psi^\dag_{\vb{k'} j}} = \delta_{i, j} \delta_{\vb{k},\vb{k'}},
\end{gather*}
where $\qty{\cdot,\cdot}$ is the anticommutator. These Nambu spinors behave like conventional electron spinor fields, but describe electron-hole pairs \cite{Coleman2015}. With this, the Hamiltonian in matrix form becomes:
\begin{gather*}
H = \sum_{\vb{k}} \psi_{\vb{k}}^\dag \vb{h}_{\vb{k}} \psi_{\vb{k}}
\\
\text{where} \quad
\vb{h}_{\vb{k}} = \vb{h}^0_{\vb{k}} + \vb*{\Sigma}_{\vb{k}}
= 
\begin{pmatrix}
\xi^{(1)}_{\vb{k}} && 0 \\
0 && \xi^{(2)}_{\vb{k}}
\end{pmatrix}
+
\begin{pmatrix}
0 && -\Delta_{\vb{k}} \\
-\Delta^*_{\vb{k}} && 0
\end{pmatrix}.
\end{gather*}
Note, how we divided the Hamiltonian into its non-interacting and self-energy part. As in Eq.~\eqref{eq:H-BCS}, the normal (diagonal) part of the self-energy $\vb*{\Sigma}$ has been absorbed by $H_0$, such that there only is an anomalous (off-diagonal) part.

\subsubsection{Bogoliubov Transformation}
Diagonalising the Hamiltonian, $\vb{h}_{\vb{k}}$ is known as a Bogoliubov transformation, see e.g. Refs. \cite{Bruus2004, Coleman2015, Fil2018}. Letting $\phi_\mathbf{k}$ denote the order parameter phase, i.e. $\Delta_\mathbf{k} = \abs{\Delta_\mathbf{k}}e^{i \phi_\mathbf{k}}$, the resulting eigen-spinors $\gamma_{\vb{k}}^\dag = \begin{pmatrix} \alpha^{(1) \dag}_{\vb{k}} && \alpha^{(2) \dag}_{\vb{k}} \end{pmatrix}$ are:
\begin{subequations}
\label{eq:bologiubov}
\begin{gather}
\label{eq:bologiubov-U}
\begin{pmatrix}
    c^{(1)}_{\vb{k}} \\[0.3em]
    c^{(2)}_{\vb{k}}
\end{pmatrix}
= \psi_{\vb{k}} = \vb{U}_{\vb{k}}
\gamma_{\vb{k}} =
\begin{pmatrix}
    u_{\vb{k}} && v^*_{\vb{k}} \\
    -v_{\vb{k}} && u^*_{\vb{k}}
\end{pmatrix}
\begin{pmatrix}
    \alpha^{(1)}_{\vb{k}} \\[0.3 em]
    \alpha^{(2)}_{\vb{k}}
\end{pmatrix},
\\[0.5em]
u_{\vb{k}}
= \sqrt{
    \frac{1}{2} \qty(
        1 + \frac{\xi_{\vb{k}}}{E_{\vb{k}}}
    )
} e^{i \phi_\mathbf{k} /2}, \quad
v_{\vb{k}}
= \sqrt{
    \frac{1}{2} \qty(
        1 - \frac{\xi_{\vb{k}}}{E_{\vb{k}}}
    )
} e^{-i \phi_\mathbf{k} /2}, \\
\label{eq:bologiubov-uk-vk}
u_{\vb{k}} v_{\vb{k}} = \frac{\abs{\Delta_{\vb{k}}}}{2 E_{\vb{k}}}, \quad u_{\vb{k}} v^*_{\vb{k}} = \frac{\Delta_{\vb{k}}}{2 E_{\vb{k}}}
\\
\label{eq:bologiubov-Ek}
E_{\vb{k}} = \sqrt{\xi_{\vb{k}}^2 + \abs{\Delta_{\vb{k}}}^2}, \\ \xi_{\vb{k}} = \frac{\xi^{e}_{\vb{k}} + \xi^{h}_{\vb{k}}}{2}, \quad \eta_{\vb{k}} = \frac{\xi^{e}_{\vb{k}} - \xi^{h}_{\vb{k}}}{2}.
\end{gather}
\end{subequations}
Here, an electron-hole transformation gives the electron, and hole dispersions: $\xi^{e}_{\vb{k}} = \xi^{(1)}_{\vb{k}}, \quad
\xi^{h}_{\vb{k}} = -\xi^{(2)}_{\vb{-k}}$. In addition, when diagonalizing the Hamiltonian the following relations were used:
\begin{gather*}
\abs{ u_{\vb{k}} }^2 + \abs{ v_{\vb{k}} }^2 = 1, \quad
\abs{ u_{\vb{k}} }^2 - \abs{ v_{\vb{k}} }^2 = \frac{ \xi_{\vb{k}} }{ E_{\vb{k}} } \\
\xi_{\vb{k}}^e = \xi_{\vb{k}} + \eta_{\vb{k}}, \quad
\xi_{\vb{k}}^h = \xi_{\vb{k}} - \eta_{\vb{k}}.
\end{gather*}
The operators $\alpha^{(1) \dag}_{\vb{k}}, \alpha^{(2) \dag}_{\vb{k}}$ represent the creation of fermionic (Bogoliubov) \acp{QP} above the condensate BCS ground state consisting of electron-hole pairs. The diagonalized Hamiltonian is:
\begin{align*}
H
=& \sum_{\vb{k}} \gamma^\dag_{\vb{k}} \qty( \vb{U}^\dag_{\vb{k}} \vb{h}_{\vb{k}} \vb{U}_{\vb{k}} ) \gamma_{\vb{k}} \\
=&
\sum_{\vb{k}}
\begin{pmatrix} \alpha^{(1) \dag}_{\vb{k}} && \alpha^{(2) \dag}_{\vb{k}} \end{pmatrix}
\begin{pmatrix}
    E_{\vb{k}} + \eta_{\vb{k}} && 0 \\
    0 && - \qty(E_{\vb{k}} - \eta_{\vb{k}})
\end{pmatrix}
\begin{pmatrix}
    \alpha^{(1)}_{\vb{k}} \\[0.3 em]
    \alpha^{(2)}_{\vb{k}}
\end{pmatrix}
\end{align*}
And the resulting \ac{QP} energies are thus:
\begin{gather}
    \label{eq:bcs_eigenenergy}
    E^\mathrm{QP}_\textbf{k} = E_{\vb{k}} \pm \eta_{\vb{k}} = \sqrt{\qty(\frac{\xi^{e}_{\vb{k}} + \xi^{h}_{\vb{k}}}{2})^2 + \abs{\Delta_{\vb{k}}}^2} \pm \frac{\xi^{e}_{\vb{k}} - \xi^{h}_{\vb{k}}}{2}.
\end{gather}
The eigenenergies for exciton superfluidity are slightly different from the eigenenergies obtained in traditional superconductors, since both the electron $\xi^{e}_{\vb{k}}$ and hole dispersion $\xi^{h}_{\vb{k}}$ occurs in the expression. However, in the simpler case where $\xi^{e}_{\vb{k}} = \xi^{h}_{\vb{k}}$ it is easy to see from \eqref{eq:bcs_eigenenergy} that it is not possible to excite Bogoliubov \acp{QP} from the BCS groundstate with energies $E_{\vb{k}}$ less than $\Delta_{\vb{k}}$. This is also evident as a gap of order $\Delta_{\vb{k}}$ in the density of states for the \acp{QP}. For a superfluid ground state to exist, we must have the gap $\Delta_{\vb{k}} > 0$. A difference in the electron and hole dispersions may therefore have severe consequences for the possibility of the formation of a superfluid state \cite{Fil2018}.

$u_{\vb{k}}, v_{\vb{k}}$ give information on the Bogoliubov \ac{QP} amplitudes, and they can be used to calculate the condensate fraction \cite{Nilsson2021}:
\begin{gather}
\label{eq:condfrac}
c = \frac{\sum_k \abs{u_k}^2 \abs{v_k}^2}{\sum_k \abs{v_k}^2}.
\end{gather}

The size of the gap can be found in a self-consistent manner by using the definition \eqref{eq:gap-1} together with \eqref{eq:bologiubov-U}:
\begin{align*}
\Delta_{\vb{k}}
= - \sum_{\vb{k'}} U^{(12)}_{\vb{k'}-\vb{k}}
u_{\vb{k'}} v^*_{\vb{k'}} \expval{1  - \alpha^{(1) \dag}_{\vb{k'}} \alpha^{(1)}_{\vb{k'}} - \alpha^{(2) \dag}_{\vb{k'}} \alpha^{(2)}_{\vb{k'}} }
\end{align*}
By using \eqref{eq:bologiubov-uk-vk} and the fact that the Bogoliubov quasiparticles are fermions, we arrive at the mean field BCS gap equation:
\begin{align*}
    \Delta_{\vb{k}} = - \sum_{\vb{k'}} & U^{(12)}_{\vb{k'}-\vb{k'}}
    \frac{\Delta_{\vb{k'}}}{2 E_{\vb{k'}}} \qty(1 - f(E_{\vb{k'}} + \eta_{\vb{k}'}) - f(E_{\vb{k'}} - \eta_{\vb{k}'}) ).
\end{align*}

\subsubsection{Matsubara Green Functions}
Following, Ref. \cite{Lozovik2009}, We now define the following Matsubara Green functions:
\begin{gather}
G^{(n n')}(\vb{k}, \tau) = - \expval{T_{\tau} c^{(n)}_{\vb{k}} (\tau) c^{(n') \dag}_{\vb{k}} (0)},
\end{gather}
where $n,n'$ still denotes the band index (or layer index, since we have a single-band model), $T_{\tau}$ is the imaginary time ordering operator and $\expval{\cdot}$ is the thermal average. Thus, $G^{(1 1)}$ is the electron propagator for layer 1, $G^{(1 2)}$ is the anomalous electron-hole Green function and ${G^{(2 2)}}^*$ is the hole propagator. The equation of motion technique (see e.g. \cite[p. 198]{Bruus2004}) immediately gives:
\begin{gather*}
- \partial_\tau G^{(n n')}(\vb{k}, \tau)
- \sum_{n''} h_{\vb{k}}^{(n n'')} G^{(n'' n)} (\vb{k}, \tau)
= \delta(\tau) \delta_{n, n'} \\
\sum_{n''}
\qty[
\qty(G^{(n n'')}_0 (\vb{k}, \tau))^{-1}
- \Sigma^{(n n'')}_{\vb{k}} ]
G^{(n'' n')} (\vb{k}, \tau)
= \delta(\tau) \delta_{n, n'},
\end{gather*}
where we defined the non-interacting Green Function:
\begin{gather*}
G^{(n n')}_0 (\vb{k}, \tau) = \qty(- \partial_\tau - \xi^{(n)}_{\vb{k}})^{-1} \delta_{n, n'}.
\end{gather*}
The equation of motion thereby becomes the familiar Dyson equation when evaluated along the imaginary frequency axis for Matsubara frequencies $i\nu_m$:
\begin{gather*}
G^{(n n')}(\vb{k}, i \nu_m) = G^{(n n')}_0(\vb{k}, i \nu_m) \\
+ \sum_{n''} G_0^{(n n)} (\vb{k}, i \nu_m) \, \Sigma^{(n n'')} (\vb{k}, i \nu_m) \, G^{(n'' n')}(\vb{k}, i \nu_m)
\end{gather*}
Or, alternatively in a Feynman diagram representation:
\begin{equation}
\label{eq:feynmann-dyson}
\includegraphics{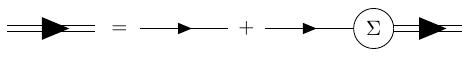}
\end{equation}
By definition, the gap in the mean-field approximation, Eq.~\eqref{eq:gap-1}, is given by the self-consistent equation:
\begin{gather}
\Delta_{\vb{k}} = \sum_{\vb{k}} U^{(12)}_{\vb{k' - k}} G^{(12)} (\vb{k'}, \tau=0^+),
\end{gather}
which in Feynman diagram notation is given by
\begin{equation}
\label{eq:feynman-self-energy}
\includegraphics{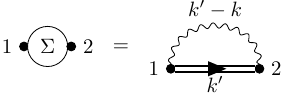}
\end{equation}

\subsubsection{Generalized Gap Equation with Screening}
So far, we have considered the simplified BCS Hamiltonian, Eq.~\eqref{eq:H-BCS}, which has a "Fock" mean field approximation to the pair interaction. The results, Eqs.~\eqref{eq:feynmann-dyson}, \eqref{eq:feynman-self-energy}, easily generalizes to the full Hamiltonian, Eq.~\eqref{eq:H-full}. The self-consistency equation, Eq.~\eqref{eq:feynman-self-energy}, shows that the BCS mean field approximation is equivalent to a simple exchange approximation to the self-energy. To treat the screening within the subspace of the model, we have instead used a statically screened exchange approximation to the self-energy in this work. The self-energy is within this approximation equivalent to the GW-approximation, calculated with the static limit of $W$:
\begin{gather}
\label{eq:gap}
\Sigma^{(nn')}_{\vb{k}} = - \sum_{\vb{k}', \sigma} W^{(n n')}_{\vb{k' - k}, \omega = 0} G^{(n n')} (\vb{k'}, \tau=0^+),
\end{gather}
where $W_{\omega = 0}$ is the static limit of the fully screened interaction. In a Feynman-diagram this reads:
\begin{equation}
 \label{eq:feynman-screened-self-energy}
\includegraphics{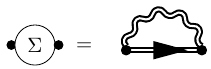}
\end{equation}
Similar to the derivation in the previous section the gap reduces to (minus) the anomalous part of the self-energy. As is shown in Section \ref{sec:cRPA}, Eq.~\eqref{eq:Wd_from_Wi}, the fully screened interaction, $W$, of the model can be determined from the effective interaction, $U$, and polarizability, $\chi_0^d$, of the model subspace of "\textbf{d}opant" bands:
\begin{gather}
\vb{W}(\omega) = \qty( \vb{1} - \vb{U}(\omega) \vb*{\chi_0}^d(\omega) )^{-1} \, \vb{U}(\omega)
\label{eq:Wgap}
\end{gather}
While this is a general equation that holds for the frequency dependent quantities we only consider the static limit of the interactions in this work. $\vb{W}$, $\vb{U}$ and $\vb*{\chi_0}^d$ are 2x2 matrices with the normal components on the diagonal and anomalous components on the off-diagonal.
We will henceforth denote the interactions by the charge carriers of each layer: $U^{(ee)}_{\vb{k}, \vb{k}'} = U^{(11)}_{\vb{k}, \vb{k}'}$, $U^{(he)}_{\vb{k}', \vb{k}} = U^{(eh)}_{\vb{k}, \vb{k}'} = U^{(12)}_{\vb{k}, -\vb{k}'}$, $U^{(hh)}_{\vb{k}, \vb{k}'} = U^{(22)}_{-\vb{k}, -\vb{k'}}$. 

In previous works \cite{pascucci2022, Nilsson2021, gupta2020,  Conti2020, Conti2019, Conti2017, Zarenia2014, Perali2013, Lozovik2012} the intrinsic screening was accounted for by a simple analytical 2D Coulomb potential for the interactions:
\begin{subequations}
\label{eq:W-analytic}
\begin{gather}
U^{(eh)}(r_\parallel) = - \frac{1}{\epsilon^i_M} \frac{1}{\sqrt{d^2 + r_\parallel^2}},
\quad
U^{(ee)}(r) = \frac{1}{\epsilon^i_M} \frac{1}{r_\parallel}.\\
U^{(eh)}(q_\parallel) = - U^{(ee)}(q_\parallel) e^{- d_\parallel q_\parallel},
\quad
U^{(ee)}(q_\parallel) = \frac{2 \pi}{\epsilon^i_M} \frac{1}{q_\parallel},
\label{eq:modelU}
\end{gather}
\end{subequations}
where $\epsilon^i_M$ has simply been set to a constant, ($\epsilon^i_M = 2$ in Refs. \cite{Conti2019, Conti2020, Nilsson2021} due to Ref. \cite{Kumar2016}). This approximation completely neglects the momentum and material dependence of the intrinsic screening. In this work we show that the effective interactions can be calculated from the \emph{ab initio} bandstructures using the \ac{QEH} framework. We will return to this in Section \ref{sec:QEH}.

In Ref. \cite{Nilsson2021}, the gap equation, Eq.~\eqref{eq:gap}, with a static exchange approximation to the self-energy was derived within a \ac{DFT} framework and an implementation of a self-consistent solver was presented (The BiEx code package). It is this code-package that we have used in this work. Similar to Ref. \cite{Conti2020}, the effect of equivalent valleys was accounted for by a simple valley degeneracy factor $g_v$. Therefore we also restrict ourselves to materials which have "direct" band gaps in the sense that the conduction band minimum and valence band maximum are located at the same high-symmetry point. Here it should be noted that since the reciprocal lattices of the two bilayers are in general different, the \ac{QP} bandgaps are only approximately direct. Further details on the implementation can be found in Ref. \cite{Nilsson2021}.

\subsection{Screening and the Random Phase Approximation \label{sec:screening}} 
The Coulomb interaction between electrons and holes in the doped bilayer system is suppressed by the polarization and the resulting screened interaction of the doped system, $W$, is a key component of the gap equation. In this section we provide an overview of linear response theory and the random phase approximation (RPA) before we derive the downfolding approximation used to compute the effective interaction $U$ in the model Hamiltonian \ref{eq:H-full} in Section \ref{sec:cRPA}. 

To determine $W$ consider the charge-charge correlation function, $\chi$. In linear response theory, $\chi$ gives the density induced by an external potential $\phi_\text{ext}(\omega)$ varying harmonically in time with frequency $\omega$:
\begin{gather}
    \rho_\text{ind}(\vb{r}, \omega) = \int\!\dd{\vb{r}'} \chi(\vb{r},\vb{r'}, \omega) \phi_\text{ext}(\vb{r}', \omega),
\end{gather}
or alternatively in matrix notation in an arbitrary basis:
\begin{gather}
\label{eq:chi-def}
    \vb*{\rho}_\text{ind}(\omega) = \vb*{\chi}(\omega) \vb*{\phi}_\text{ext}(\omega).
\end{gather}
As external field we take the instantaneous Coulomb potential from a point charge located at a position $\vb{r}'$:
\begin{gather}
\label{eq:V}
\phi_\text{ext}(\vb{r}, \omega) = V(\vb{r},\vb{r}') \equiv {\abs{\vb{r} - \vb{r}'}}^{-1},
\end{gather}
where we use atomic units in the last step.
We insert Eq.~\eqref{eq:V} into Eq.~\eqref{eq:chi-def} and introduce the dielectric function $\epsilon$ and the non-interacting polarizability $\chi_0$ through the defining relations:
\begin{subequations}
\label{eq:chi-P-eps-def}
\begin{align}
\vb*{\rho}_\text{ind}(\omega) &= \vb*{\chi}(\omega) \vb{V} \\ \vb*{\rho}_\text{ind}(\omega) &= \vb*{\chi_0}(\omega) \vb{W}(\omega) \\
\label{eq:eps-d}
\vb{W}(\omega) &= \vb*{\epsilon}^{-1}(\omega) \vb{V} \\
\vb{W}(\omega) &= \vb{V} + \vb{V} \vb*{\rho}_\text{ind}(\omega).
\end{align}
\end{subequations}
From the definitions in Eq.~\eqref{eq:chi-P-eps-def} follows the Dyson equations:
\begin{subequations}
\begin{align}
\label{eq:P-Dyson}
\vb{W}(\omega) &= \vb{V} + \vb{V} \, \vb*{\chi_0}(\omega) \, \vb{W}(\omega) \\
\vb*{\chi}(\omega) &= \vb*{\chi_0}(\omega) + \vb*{\chi_0}(\omega) \vb{V} \vb*{\chi}(\omega).
\end{align}
\end{subequations}
The lowest order polarization process is indicated on Fig.~\ref{fig:bilayer} by the Feynman pair bubble which first creates and later annihilates an electron-hole pair somewhere on the bands. In the \ac{RPA}, $\chi_0$ is calculated as the non-interacting polarizability by summing up all such transitions between occupied and unoccupied states of the doped bilayer system \cite{Hybertsen1987, Aryasetiawan1998, Aryasetiawan2022}:
\begin{align}
&\vb*{\chi_0}\qty(\vb{r}, \vb{r}', \omega) =
\sum_{n, n'}
\sum_{\vb{k}, \vb{q}}^{\mathrm{BZ}}
\qty(f_{n \vb{k}}-f_{n' \vb{k}+\vb{q}}) \alpha_{n \vb{k}, n' \vb{k} + \vb{q}}(\vb{r}, \vb{r}'),
\\
\label{P-RPA}
&\alpha_{n \vb{k}, n' \vb{k} + \vb{q}}(\vb{r}, \vb{r}')
\equiv \nonumber \\
&\frac{\phi_{n \vb{k}}^*(\vb{r}) \phi_{n' \vb{k}+\vb{q}}(\vb{r}) \phi_{n \vb{k}}\qty(\vb{r}') \phi_{n' \vb{k}+\vb{q}}^*\qty(\vb{r}')}{\omega+\varepsilon_{n \vb{k}}-\varepsilon_{n' \vb{k}+\vb{q}} +i 0^+\mathrm{sgn}(\varepsilon_{n \vb{k}}-\varepsilon_{n' \vb{k}+\vb{q}})}.
\end{align}
Here $f_{n \vb{k}}, \phi_{n \vb{k}}, \varepsilon_{n \vb{k}}$ are respectively the occupation number (in the doped system), eigenfunction and eigenenergy of band $n$ with wavevector $\vb{k}$ lying within the first Brillouin zone. In the superfluid phase the normal polarization in Eq.~\eqref{P-RPA} is supplemented by a corresponding anomalous component that reflects the quenching of screening in the superfluid phase. In this work we compute the real and anomalous model polarization within the RPA, following Refs. \cite{lozovik1997, Lozovik2012, Conti2019} using the implementation from Ref. \cite{Nilsson2021} (see discussion below). 

\subsection{Downfolding of Screening \label{sec:cRPA}}
The screening in the superfluid phase is determined self-consistently by solving the gap equation \ref{eq:gap}-\ref{eq:Wgap}. As discussed in Section \ref{sec:gap} the gap equation is restricted to the subspace spanned by the dopant bands (Fig.~\ref{fig:bilayer}). A crucial parameter for the gap equation is the effective interaction $U$ of the model Hamiltonian (Eq.~\ref{eq:H-full}). $U$ can be defined by requiring that the fully screened interaction of the model Hamiltonian in Eq.~\ref{eq:H-full} coincides with the fully screened interaction of the full Hamiltonian. As we will see below this implies that $U$ is partially screened by all screening channels not explicitly included in the low energy model. In this Section we show that $U$ for the doped system can be identified as the fully screened interaction of the undoped system.
\\

We will determine the polarization $\chi_0$ using the downfolding strategy of the \ac{cRPA} \cite{Aryasetiawan2004, Aryasetiawan2022}. As indicated in Fig.~\ref{fig:bilayer}, the core idea is to divide the bands into the \emph{dopant} bands, $\mathcal{D}$, i.e. the partially occupied conduction and valence band for the n- and p-doped layer respectively, and the remaining \emph{intrinsic} bands. Hereby we can distinguish between intraband (metallic) screening processes such as the orange pair bubble in Fig.~\ref{fig:bilayer} and interband (intrinsic) processes such as the black pair bubble. At low temperatures for intrinsic (i.e. undoped) layers there are no electrons or holes in the dopant bands. Thus, the intrinsic system has no metallic screening. We therefore decompose the polarization $\chi_0$ into a metallic, $\chi_0^d$ ($d$ for doped), and an intrinsic, $\chi_0^i$, polarization:
\begin{gather}
\label{eq:Pi-downfolding}
\vb*{\chi_0}(\omega) \approx \vb*{\chi_0}^d(\omega) + \vb*{\chi_0}^i(\omega),
\end{gather}
$\chi_0^i$ includes all intrinsic transitions (between occupied and unoccupied bands of the undoped system):
\begin{align*}
\chi_0^i\qty(\vb{r}, \vb{r}', \omega)=
\sum_{n, n'}
\sum_{\vb{k}, \vb{q}}^{\mathrm{BZ}}
\qty(f_{n \vb{k}}^i-f_{n' \vb{k}+\vb{q}}^i) \alpha_{n \vb{k}, n' \vb{k} + \vb{q}}(\vb{r}, \vb{r}').
\end{align*}
Here we let $f^i_{n \vb{k}}$ denote the intrinsic occupation number which at zero temperature is simply 0 for energies below the intrinsic Fermi-level and 1 for energies above. Meanwhile in our downfolding approximation $\chi_0^d$ is restricted to states in the subspace of the dopant bands ($n,n' \in \mathcal{D}$). The normal (diagonal) components of the polarization is given by:
\begin{align}
\label{eq:Pi-d}
\chi_0^d \qty(\vb{r}, \vb{r}', \omega)=
\sum_{n, n' \in \mathcal{D}}\sum_{\vb{k}, \vb{q}}^{\mathrm{BZ}}
\qty(f_{n \vb{k}}-f_{n' \vb{k}+\vb{q}}) \alpha_{n \vb{k}, n' \vb{k} + \vb{q}}(\vb{r}, \vb{r}')
\end{align}
In the superfluid state $\chi_0^d$ is supplemented by the anomalous polarization which comes in as off-diagonal components in the 2x2 representation in Eq.~\eqref{eq:Wgap}. $\chi_0^d$ is calculated self-consistently together with the anomalous polarization when solving the gap equation, Eq.~\eqref{eq:gap}, for the downfolded Hamiltonian of Eq.~\eqref{eq:H-full} (spanned only by dopant bands). By comparison of $\chi_0, \chi_0^i, \chi_0^d$ we see that the downfolding approximation, Eq.~\eqref{eq:Pi-downfolding}, neglects the screening channel $\chi_0^{di}$ coming from transitions between the dopant bands and higher order intrinsic bands. Since we assume low doping concentrations this simplification should be justifiable.

Isolating $\vb{W}$ in the Dyson equation \eqref{eq:P-Dyson} gives:
\begin{gather}
\label{eq:Wd}
\vb{W}(\omega) = (\vb{1} - \vb{V} \vb*{\chi_0}(\omega))^{-1} \vb{V}
\end{gather}
This motivates us to define the intrinsic screened interaction, $U$, which only takes into account the intrinsic polarizability $\chi_0^i$:
\begin{gather}
\vb{U}(\omega) \equiv (\vb{1} - \vb{V} \vb*{\chi_0}^i(\omega))^{-1} \vb{V} \label{eq:Wi}
\end{gather}
It is no coincidence, that we name the intrinsic screened interaction, $U$. As we shall soon see, it acts as the effective interaction of the downfolded model, Eq.~\eqref{eq:H-full}.
Isolation of $\chi_0$ and $\chi_0^i$ in \eqref{eq:P-Dyson} and \eqref{eq:Wi} gives:
\begin{gather}
\notag
\left.
\begin{array}{l}
    \vb*{\chi_0}(\omega) = \qty(\vb{V})^{-1} - \qty(\vb{W}(\omega))^{-1} \\
    \vb*{\chi_0}^i(\omega) = \qty(\vb{V})^{-1} - \qty(\vb{U}(\omega))^{-1}
\end{array}
\right\}
\Rightarrow \\
\label{eq:Wd_from_Wi}
\vb{W}(\omega) = \vb{U}(\omega) + \vb{U}(\omega) \vb*{\chi_0}^d(\omega) \vb{W}^d(\omega)
\end{gather}
This is yet another Dyson equation and by comparison with Eq.~\eqref{eq:P-Dyson} it shows the intrinsic screened interaction $U$, which is the fully screened interaction of the undoped system, acts as the "effective bare Coulomb interaction" in the subspace of the dopant bands. We will henceforth stick to the following terminology:
\begin{align*}
W &= \text{"The doped screened interaction"}. \quad \text{(Eq.~\eqref{eq:Wd})}\\
U &= \text{"The intrinsic screened interaction"}. \quad \text{(Eq.~\eqref{eq:Wi})} \\
V &= \text{"The bare Coulomb interaction"}. \quad \text{(Eq.~\eqref{eq:V})}
\end{align*}
Since the intrinsic screened interaction $U$ includes all polarization processes for the undoped bilayer structure it is simply calculated from the dielectric response $\chi^i$ of the intrinsic heterostructure (Eq.~\eqref{eq:P-Dyson}):
$$
\vb{U}(\omega) = \qty(\vb{1} + \vb{V} \vb*{\chi}^i(\omega)) \vb{V}
$$
As explained in Section \ref{sec:QEH}, we determine the intrinsic response $\chi^i$ through the \ac{QEH} framework.

\subsection{The \ac{QEH} Framework \label{sec:QEH}}
The intrinsic screened interaction, $U$, of the entire \ac{vdWH} is calculated from first principles using the \ac{QEH} model \cite{Andersen2015, Latini2015, Latini2017}. An open-source Python implementation of the \ac{QEH} model interfaced with the GPAW electronic structure code \cite{GPAW_orig, GPAW_review} is found at Ref. \cite{QEH_documentation}. At its core, the \ac{QEH} model utilizes that the weak interactions between layers in a \ac{vdWH} means that hybridization between layers is small. In other words it is assumed that the electron states are well localized in the 2D-layers such that there is little overlap of wavefunctions between layers. The properties of the entire \ac{vdWH} is then obtained by combining \ac{DFT}-calculations of individual layers via
electrostatic interactions \cite{Andersen2015}. The general objective is to calculate the dielectric function $\vb*{\epsilon}^i$ of the undoped \ac{vdWH}. From this, we obtain the effective screened potential for the undoped system using an equation similar to \eqref{eq:eps-d}, but for the intrinsic screened interaction $U$. Summarised, the approach of \ac{QEH} is:
\begin{enumerate}
    \item For each layer: From DFT, calculate in plane-averaged response function and induced density $(\vb*{\chi}^i_\parallel, \rho_\text{ind})$.
    \item Then calculate the density response $\chi^i$ of the entire vdWH by solving a Dyson equation that couples $(\vb*{\chi}^i_\parallel, \rho_\text{ind})$ together via Coulomb interaction.
\end{enumerate}
For a detailed discussion of the \ac{QEH} framework see \cite{Andersen2015, Latini2015}.

\begin{figure*}[ht!]
\centering
\begin{subfigure}[t]{0.495\textwidth}
    \caption{Intrinsic static macroscopic dielectric function $\epsilon^i_M$}
    \includegraphics[width=\textwidth]{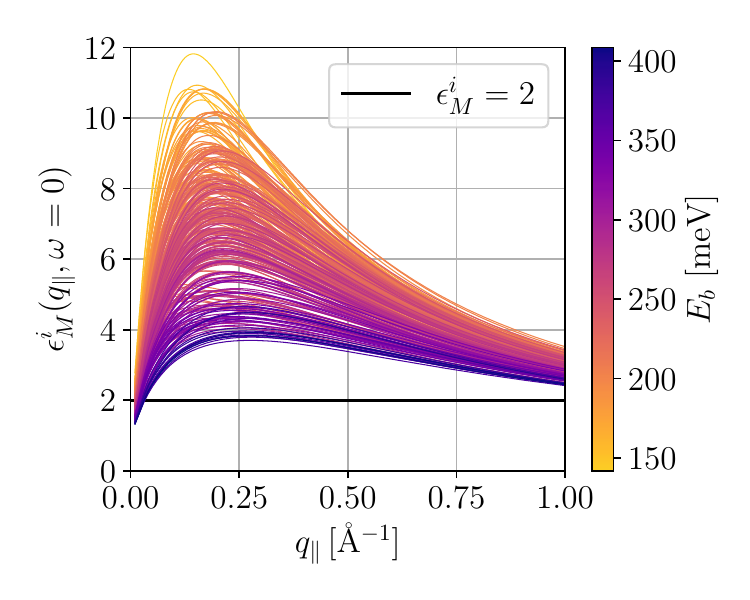}
    \label{fig:epsM}
\end{subfigure}
\hfill
\begin{subfigure}[t]{0.495\textwidth}
    \caption{Screened electron-hole potential $U^{(eh)}$}
    \includegraphics[width=\textwidth]{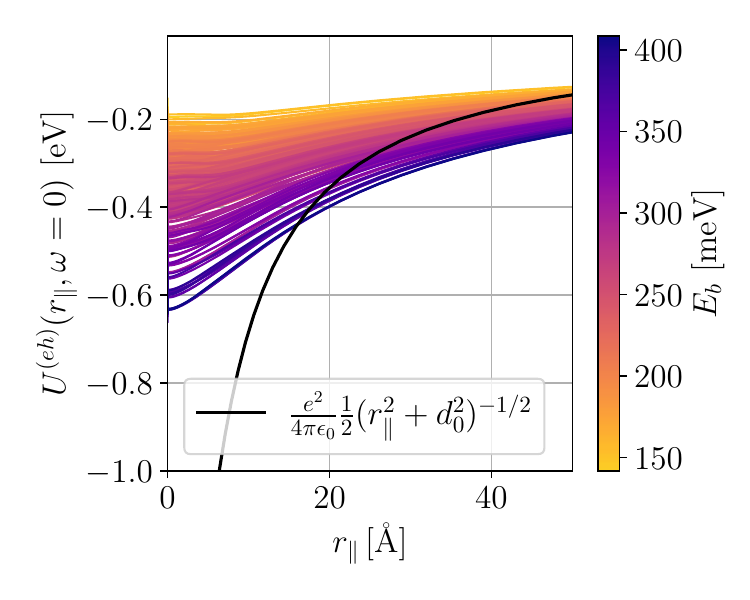}
    \label{fig:W_i_r}
\end{subfigure}
\caption{Intrinsically screened macroscopic dielectric function $\epsilon^i_M$ as a function of the in-plane momentum transfer (left) and intrinsic static screened electron-hole interaction strength $U^{(eh)}$ as a function of the in-plane distance (right) calculated through the \emph{ab initio} \ac{QEH} method for all 336 \ac{TMD}/\ac{TMO} bilayer combinations. For comparison the analytic result with fixed $\epsilon^i_M=2$ as used in \cite{Conti2020, Nilsson2021} is also shown in black on both plots. As indicated on the color bar, the line colors of the coloured graphs correspond to the exciton binding energies $E_b$ given by the Mott-Wannier model, Eq. \eqref{eq:Mott-Wannier}.}
\label{fig:epsM_and_W_i_r}
\end{figure*}

\subsection{Mott-Wannier Model \label{sec:Mott-Wannier}}
A simple model of the electron-hole pairs is provided by the Mott-Wannier model \cite{Wannier1937} which reduces the full many-body problem to a simplified hydrogenic model where the electron and hole interact via a screened Coulomb interaction. A generalization to 2D systems is derived in Refs. \cite{Latini2015} (also see Ref. \cite{Andersen2015}) and gives the hydrogenic eigenvalue problem:
\begin{gather}
\label{eq:Mott-Wannier}
\qty(\frac{- \grad_\text{2D}^2}{2 m_\text{ex}} + U^{(eh)}(\vb{r}_\parallel)) F_n(\vb{r}_\parallel) = E_n F_n(\vb{r}_\parallel),
\end{gather}
where $E_n$ is the $n$'th eigenvalue, $F_n(\vb{r})$ is the $n$'th exciton wavefunction, $m_\text{ex} \equiv (1/m_e + 1/m_h)^{-1}$ is the effective mass and $U^{(eh)}$ is the intrinsic screened interaction between the spatially separated electrons and holes. The exciton binding energy is defined from the ground state energy: $E_b = - E_1$. "Classically", we recognize that for $E_b = - E_1$ to be large such that we have strong binding, we want a small kinetic energy (and therefore a big exciton mass) and a strong interaction $U^{(eh)}(\vb{r})$ (i.e. as negative in position space as possible). On physical grounds one would expect that the exciton condensation is correlated to how strongly the excitons are bound in the hydrogenic Mott-Wannier orbitals. It is clear that in the limit $E_b \to 0$, superfluidity vanishes and it can be shown that high exciton binding energy is a requirement for multi-component superfluidity (multicomponent due to spin-orbit coupling) \cite{Conti2020}. Therefore it is of interest to investigate the $E_b$-values. However, as is discussed in Ref. \cite{Nilsson2021}, the correlation between $E_b$ and the superfluid gap $\Delta$ is not strict and the full gap equation, Eq.~\eqref{eq:gap} is needed to properly characterise the superfluid properties of the \acp{vdWH}. 
\begin{figure*}[ht!]
\centering
\begin{subfigure}[t]{0.497\textwidth}
    \caption{$E_b$ calculated with fixed $\epsilon^i_M = 2$ and fixed interlayer distance of $d_0 = 9.1$ Å (average $d$ of all \acp{vdWH}).}
    \includegraphics[width=\textwidth]{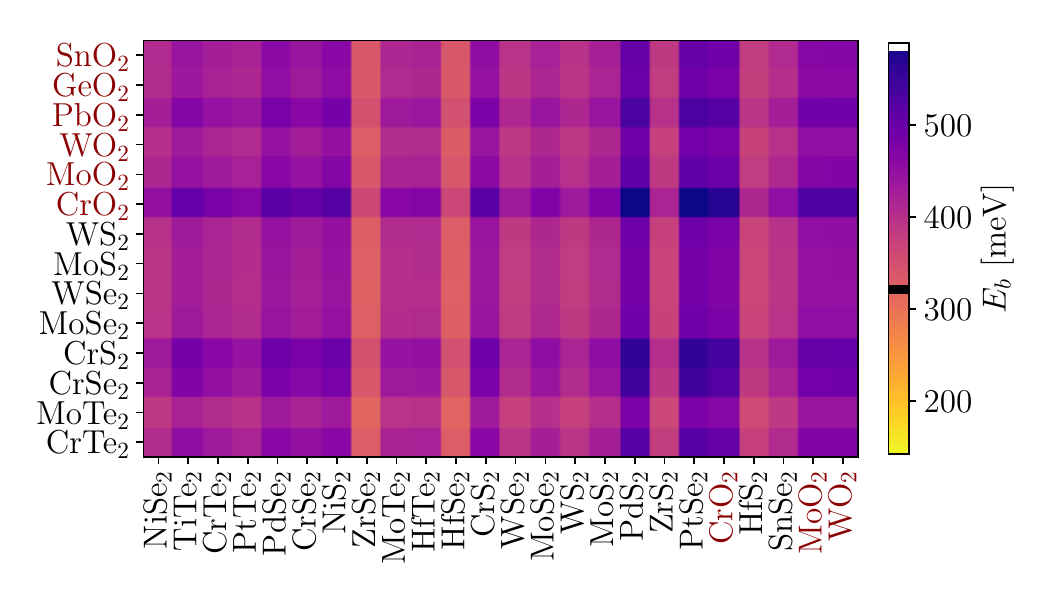}
    \label{fig:E_b_analytic_const_d}
\end{subfigure}
\hfill
\begin{subfigure}[t]{0.497\textwidth}
    \caption{$E_b$ calculated with fixed $\epsilon^i_M = 2$ but variable distance $d$.}
    \includegraphics[width=\textwidth]{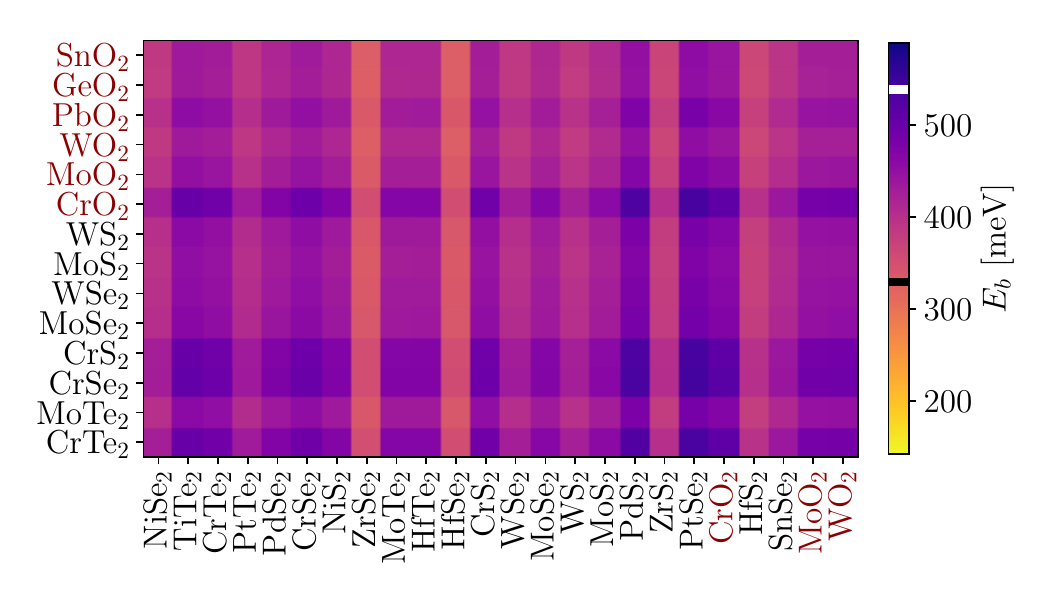}
    \label{fig:E_b_analytic}
\end{subfigure}
\hfill
\begin{subfigure}[t]{0.99\textwidth}
    \caption{$E_b$ with the screened interaction calculated using the ab initio \ac{QEH} method.}
     \includegraphics[width=\textwidth]{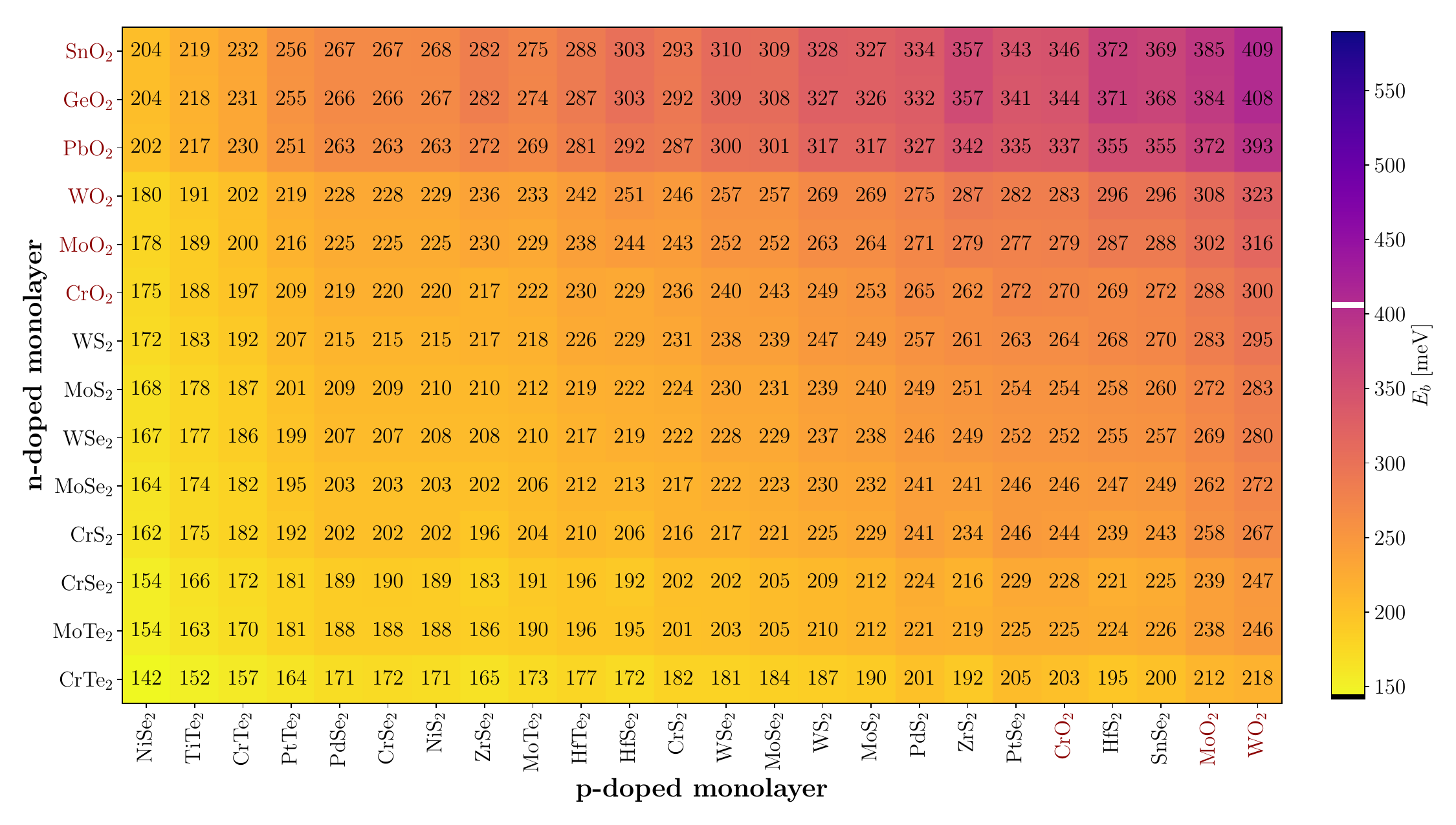}
    \label{fig:E_b_ab_initio}
\end{subfigure}
\caption{Heatmaps of interlayer exciton binding energies $E_b$ for the 336 different \ac{TMD}/\ac{TMO} bilayer combinations using 3 different methods of calculation. All plots share the same $E_b$-scale and ordering of monolayers so that colors and patterns herein can be compared. The sorting is determined by the \emph{ab initio} calculations such that the lower left corner has the weakest $E_b$-values, while the upper right corner has the strongest. The minimal and maximal $E_b$-values for each heatmap are indicated by the respectively black and white lines on the colorbars. \Acp{TMO} are written in dark red and \acp{TMD} are written in black.}
\label{fig:E_b}
\end{figure*}

\section{Results \label{sec:results}}
A comparison between the \emph{ab initio} screened interaction computed using the QEH framework (see Section \ref{sec:cRPA} - \ref{sec:QEH}) with the commonly used fixed-$\epsilon^i_M$ model interaction  (Eq: \ref{eq:W-analytic}) is shown in Fig.~\ref{fig:epsM_and_W_i_r} for the 336 \ac{TMD}/\ac{TMO} bilayer combinations. In Fig.~\ref{fig:epsM} we show the static dielectric function $\epsilon^i_M(q_\parallel, \omega=0)$ in (in-plane) momentum space while Fig.~\ref{fig:W_i_r} displays the static screened potential $U^{(eh)}(r_\parallel, \omega=0)$ between electrons and holes in (in-plane) position space. Note, that when computing $U$ using the \ac{QEH} we consider the complete dielectric function in the monopole/dipole basis (and not just the macroscopic dielectric function shown in Fig.~\ref{fig:epsM}). The color of each line in both Fig.~\ref{fig:epsM} and \ref{fig:W_i_r} reflects the interlayer exciton binding energy which was calculated using the \ac{QEH} framework through the Mott-Wannier model, Eq.~\eqref{eq:Mott-Wannier} (see Ref. \cite{Andersen2015} for details). As expected, the q-independent permittivity $\epsilon^i_M=2$ which is also used in Refs. \cite{Kumar2016, Conti2020, Nilsson2021} clearly underestimates the intrinsic screening and completely neglects the variation in screening across the different heterostructures. The maximum values of the momentum-dependent dielectric functions range from $\sim 4$ to $\sim 12$. The differences in screening is also reflected in the electron-hole potentials in Fig.~\ref{fig:W_i_r}, where we have included the analytic model interaction from Eq.~\eqref{eq:W-analytic} calculated with fixed $\epsilon^i_M=2$. The potential-well of the analytic interaction is clearly deeper and more narrow than the \emph{ab initio} potential wells. This again underlines that screening in 2D is not well-described by a single q-independent dielectric constant. Again we find significant material dependent variation in the interaction strength, which is ignored in the analytical model. As such the only material-dependent parameter in the analytical Coulomb-interaction (Eq.~\eqref{eq:W-analytic}) is the interlayer distance $d$.

The screened interaction $U$ is calculated from the full dielectric tensor, but nevertheless, we expect the macroscopic dielectric function to be an indicator of the amount of screening. From Eq.~\eqref{eq:Mott-Wannier} we would expect that a stronger interaction (from less screening) gives a stronger exciton binding energy. A look at the coloring on both Fig.~\ref{fig:epsM} and \ref{fig:W_i_r} shows that this is indeed the general trend: Weaker interaction gives weaker binding while stronger interaction gives stronger binding. However, since the effective mass also comes into play as a material dependent parameter in Eq.~\eqref{eq:Mott-Wannier}, the trend is not strictly satisfied. Some materials have large exciton binding energies even though they are not the least screened. A final point which is illustrated in Ref. \cite{Andersen2015}, but not evident from Fig.~\ref{fig:epsM_and_W_i_r} is that the bandgaps of the \acp{TMD} and \acp{TMO} constituting the \ac{vdWH} are also correlated with the screening since larger intralayer-bandgaps reduce the screening \cite{Andersen2015}. Therefore a larger band gap should also mean higher exciton binding energies and potentially higher cutoff densities and transition temperatures to the superfluid state \cite{Conti2020}.

\subsection{Exciton Binding Energy}
We now continue to discuss the exciton binding energy. In Fig.~\ref{fig:E_b} we show three heatmaps of the binding energies for the 336 \ac{TMD}/\ac{TMO} bilayer combinations using three different methods of calculations. All plots have the same scale on the color bar and ordering of monolayer labels so that patterns between the plots can be compared. The first heatmap, Fig.~\ref{fig:E_b_analytic_const_d}, shows the results obtained using the analytical Coulomb interaction from Eq.~\eqref{eq:W-analytic} with a fixed $\epsilon^i_M=2$ and fixed interlayer distance $d_0 = 9.1 \, \text{Å}$. This distance was chosen as the average interlayer distance for all material combinations. The next heatmap, Fig.~\ref{fig:E_b_analytic}, shows the same results but now with variable interlayer distance (calculated by relaxation of the homobilayers as explained in Section \ref{sec:selecting-materials}). Finally, the large heatmap in Fig.~\ref{fig:E_b_ab_initio} shows the exciton binding energies obtained using the screened interactions $U(q_\parallel, \omega=0)$ (from Fig.~\ref{fig:epsM_and_W_i_r}) calculated using the \emph{ab initio} \ac{QEH} method. As our first observation we notice that the exciton binding in the \emph{ab initio} model is significantly weaker than in the model using the analytic screened Coulomb interaction from Eq.~\ref{eq:W-analytic} with fixed permittivity. The binding energy range lies between about 350 meV - 540 meV for the analytical models, but between 140 meV - 410 meV for the \emph{ab initio} model. This aligns well with the results in Fig.~\ref{fig:epsM_and_W_i_r} where we showed that the \emph{ab initio} model had significantly more screening, thus giving a weaker electron-hole interaction resulting in lower $E_b$-values. 

The next immediate observation from comparing the heatmaps in Figs.~\ref{fig:E_b_analytic_const_d}, \ref{fig:E_b_analytic} and \ref{fig:E_b_ab_initio} is that the patterns in the variation of $E_b$ between materials are different in the different approximation schemes. The \emph{ab initio} results in Fig. \ref{fig:E_b_ab_initio} suggest that the materials in the top right corner, all including TMOs in the $n$-doped layer, have the highest exciton binding energies.
The reason for this is the large band gaps of these materials which give a much smaller screening and thus larger electron-hole interactions and exciton binding energies. This trend is clearly not captured by the analytic model interaction with fixed $\epsilon^i_M$ in Figs.~\ref{fig:E_b_analytic_const_d}, \ref{fig:E_b_analytic}, since the material dependent screening is ignored in these calculations.

In Fig.~\ref{fig:E_b_analytic_const_d} it is only the effective exciton mass which determines the variation in $E_b$ since $\epsilon^i_M, d$ are fixed. In other words,  Fig.~\ref{fig:E_b_analytic_const_d} is essentially a way of showing which \acp{vdWH} have the largest effective masses. In Fig.~\ref{fig:E_b_analytic} the interlayer distance is also allowed to vary and this changes the pattern slightly, but it still does not capture the \emph{ab initio} trend of bilayers with n-doped oxides having larger exciton binding energies. Thus to accurately describe the effect of screening on the exciton binding and evaluate trends across different materials we need the more accurate \emph{ab initio} methods.
 
\subsection{Superfluid gap and condensate fraction \label{sec:superfluid-gap-results}}
Though the exciton binding energies give some information on the superfluid properties we need to consider the results from the full gap equation, Eq.~\eqref{eq:gap}, to also see the effect of the like-particle interactions $U^{(ee)}, U^{(hh)}$ and the competition between the normal metallic intraband screening from $\chi_0^d$ and the anomalous counterpart. 

\begin{figure*}[ht!]
\centering
\textbf{Superfluidity results with fixed }$\epsilon^i_M = 2$ \par\medskip
\begin{subfigure}[t]{0.625\textwidth}
    \caption{Maximal superfluid gap versus density}
    \includegraphics[width=\textwidth]
    {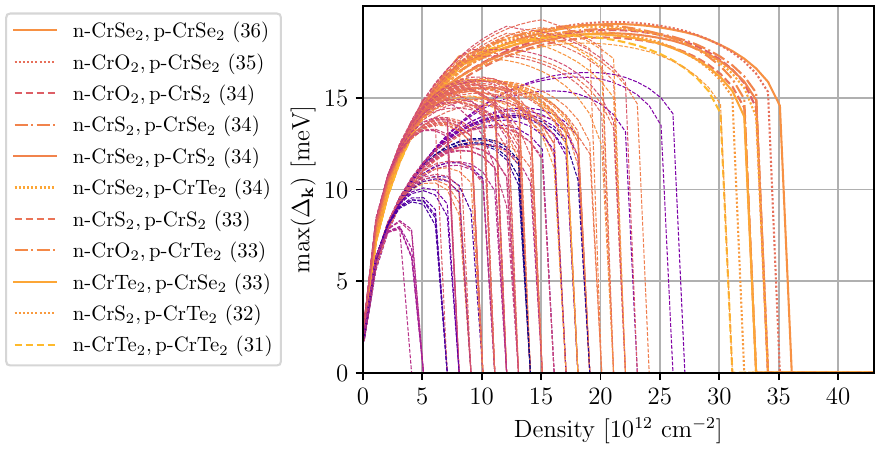}
    \label{fig:deltamax-analytic}
\end{subfigure}
\hfill
\begin{subfigure}[t]{0.368\textwidth}
    \caption{Condensate fraction versus density.}
    \includegraphics[width=\textwidth]{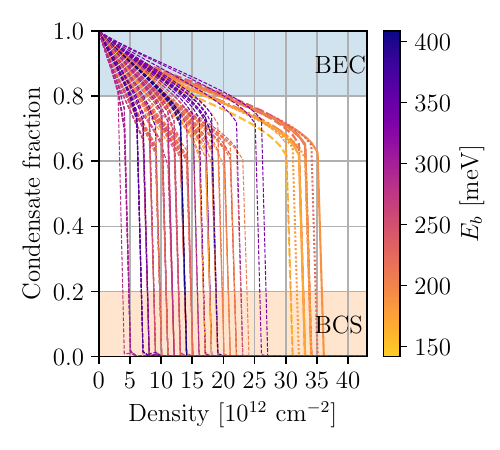}
    \label{fig:condfrac-analytic}
\end{subfigure}
\caption{Results from solving the gap equation for the bilayer system while accounting for the static screening by assuming a fixed $\epsilon^i_M = 2$. The bilayers have been sorted by their cutt-off densities and the top 10 structures are highlighted in the legend together with their cut-off densities in units of $10^{12} \, \mathrm{cm^{-2}}$ (written in parentheses). Similar to Figure \ref{fig:epsM_and_W_i_r} and \ref{fig:E_b} the lines are colored by their exciton-binding energy values (i.e. $E_b$-values from ab initio method as seen on Figure \ref{fig:E_b_ab_initio}).}
\label{fig:gap-analytic}
\end{figure*}

\begin{figure*}[ht!]
\centering
\textbf{Superfluidity results with ab initio interaction}
\begin{subfigure}[t]{0.625\textwidth}
    \caption{Maximal superfluid gap versus density.}
    \includegraphics[width=\textwidth]{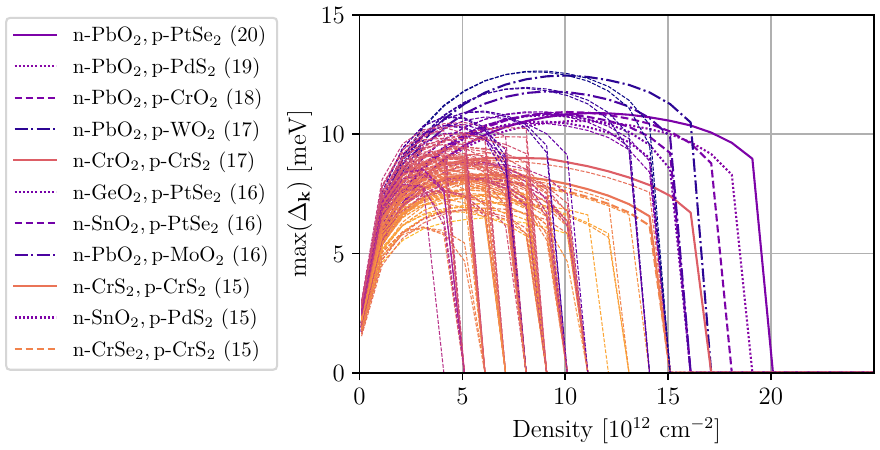}
    \label{fig:deltamax-qeh}
\end{subfigure}
\hfill
\begin{subfigure}[t]{0.368\textwidth}
    \caption{Condensate fraction versus density}
    \includegraphics[width=\textwidth]{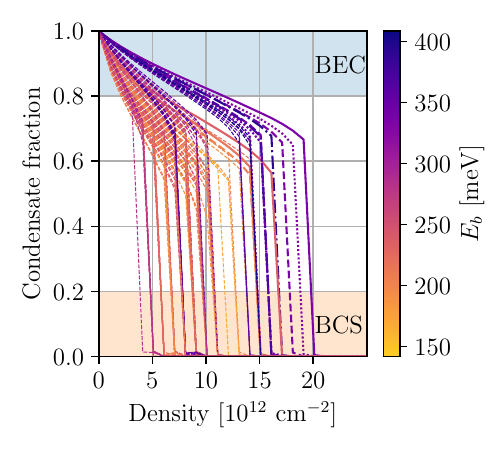}
    \label{fig:condfrac-qeh}
\end{subfigure}
\caption{Results from solving the gap equation for the bilayer system using ab initio calculations to account for the static screening. The bilayers have been sorted by their cut-off densities and the top 10 structures are highlighted in the legend together with their cut-off densities in units of $10^{12} \, \mathrm{cm^{-2}}$ (written in parentheses). Similar to Figure \ref{fig:epsM_and_W_i_r} and \ref{fig:E_b} the lines are colored by their exciton-binding energy values (i.e. $E_b$-values from ab initio method as seen on Figure \ref{fig:E_b_ab_initio}).}
\label{fig:gap-qeh}
\end{figure*}

In Fig.~\ref{fig:gap-analytic} we present the gap equation results with a fixed intrinsic permittivity $\epsilon^i_M=2$ and in Fig.~\ref{fig:gap-qeh} the same results with \emph{ab initio} intrinsic screening. As explained in Section \ref{sec:gap}, we have only considered materials with direct interlayer bandgaps in our superfluidity calculations, which means that only 136 out of the 336 different bilayer combinations are included in this final part of the study.

Fig.~\ref{fig:deltamax-analytic} and \ref{fig:deltamax-qeh} show the maximum values of the $\mathbf{k}$-dependent superfluid gap in Eq.~\eqref{eq:gap}, ($\Delta = \mathrm{max}(\Delta_\mathbf{k})$)  for different particle densities while Fig.~\ref{fig:condfrac-analytic} and \ref{fig:condfrac-qeh} show the condensate fractions, Eq.~\eqref{eq:condfrac}. The in-plane densities are the number of electrons/holes per area and they are assumed equal for the two layers. We measure the "superfluidity performance" of the bilayers with the cutoff density, $n_\text{cutoff}$, which is the maximal density for which $\Delta$ is positive. In other words, the cutoff density is the density beyond which superfluidity is killed \cite{Nilsson2021}. 
In 2D the superfluid transition is of Berezinskii-Kosterlitz-Thouless type \cite{kosterlitz1973}, and the corresponding transition temperature ($T^{\mathrm{BKT}}$) can be estimated 
from the cutoff density \cite{saberi2020}, $n_\text{cutoff}$, as 
\begin{align}
    T^{\mathrm{BKT}} = \frac{\pi n_\text{cutoff} \hbar^2}{2g_s g_v (m_e^* + m_h^*)},
    \label{TKTeq}
\end{align}
where $g_s$ and $g_v$ are the spin and valley degeneracy factors. Thus, high cutoff densities also imply high transition temperatures.
It should be noted that we neglect the frequency dependence of the dynamical screening in all calculations. A suggestion for treating the dynamical screening in a consistent manner is found in Ref. \cite{Nilsson2021}, where the authors show that neglecting the frequency dependence yields a systematic underestimation of the cut-off densities. However, this treatment is not feasible for our large-scale calculations and therefore our calculations give lower limits to the cut-off densities.

In the legends of both Fig.~\ref{fig:gap-analytic} and \ref{fig:gap-qeh} we have only labeled the top 10 performers in terms of their $n_\text{cutoff}$ values. Therefore the materials labeled in Fig.~\ref{fig:gap-analytic} are not the same as those labeled in Fig.~\ref{fig:gap-qeh}. In fact there is little overlap between the top performers in the two figures which shows that the \emph{ab initio} model gives different results compared to the fixed-$\epsilon^i_M$ model. This point is further emphasized by the fact that the top-performing structures of the \emph{ab initio} results, Fig.~\ref{fig:gap-qeh}, have significantly lower cut-off densities (15-20 $\cdot 10^{12}$ $\mathrm{cm^{-2}}$) compared to the fixed-$\epsilon^i_M$ densities (31-36 $\cdot 10^{12}$ $\mathrm{cm^{-2}}$) on Fig.~\ref{fig:gap-analytic}. This is in accordance with our previous analysis on screening and binding energies which showed that the analytical fixed-$\epsilon^i_M$ model underestimates the detrimental effect of screening and thereby overestimates the electron-hole interaction strength.

Both with fixed $\epsilon^i_M$ and \emph{ab initio} interaction, the condensate does not reach the BCS regime. Instead we see both in the density and condensate fraction an abrupt decline at the cut-off. This is a characteristic feature for static mean-field treatments \cite{Nilsson2021, Conti2020}. In Ref. \cite{Nilsson2021}, this discrete jump was found to disappear when accounting for dynamic screening in double bilayer graphene, and it was shown that the condensate can actually be tuned to the BCS-regime for these platforms. One can speculate that a full treatment of the frequency dependence could yield similar conclusions for our compounds. However, the effective treatment of the frequency dependence suggested in Ref. \cite{Nilsson2021} is relatively involved and not feasible for high throughput studies like this one.

Finally, we present in Fig.~\ref{fig:n_cutoff_heatmap_analytic} and \ref{fig:n_cutoff_heatmap_QEH} heatmaps for the cutoff densities (corresponding to Fig.~\ref{fig:gap-analytic} and \ref{fig:gap-qeh}, respectively), obtained by solving the gap equation, Eq.~\eqref{eq:gap}, with $\epsilon_M^i=2$ and \emph{ab initio} input parameters, respectively. The blank spaces in the heatmaps indicate the 200 materials with indirect bandgaps which have been omitted from the gap equation calculations as explained in Section \ref{sec:gap}. The materials in Fig.~\ref{fig:n_cutoff_heatmap} are listed in the same order as in the previous figures (Fig.~\ref{fig:E_b}) so that we can compare patterns between the figures. In other words the upper right corner in Fig.~\ref{fig:n_cutoff_heatmap} are the materials with the highest exciton binding energies using the \emph{ab initio} interaction. Fig.~\ref{fig:n_cutoff_heatmap_analytic} and \ref{fig:n_cutoff_heatmap_QEH} naturally emphasize the points made from Fig.~\ref{fig:gap-analytic} and \ref{fig:gap-qeh}: that there are different top performers in the two methods, and that the \emph{ab initio} top performers have much lower cut-off densities than the fixed-$\epsilon_M^i$ top performers. 
In Fig.~\ref{fig:n_cutoff_heatmap_analytic}, one might notice a reminiscence of the patterns in the fixed-$\epsilon_M^i$ heatmaps in Figs.~\ref{fig:E_b_analytic_const_d} and \ref{fig:E_b_analytic} (showing the binding energies using constant interlayer distance $d$ and variable $d$, respectively). From a physical perspective this indicates that the pattern in  Fig.~\ref{fig:n_cutoff_heatmap_analytic} follows the exciton masses of each bilayer. This is illustrated by the very clear (linear) correlation between the fixed-$\epsilon_M^i$ cutoff density (on the 2nd axis) and exciton mass (on the 1st axis) displayed in the scatterplot of \ref{fig:n_vs_m_analytic}.
We observe that in going from Fig.~\ref{fig:n_cutoff_heatmap_analytic} to \ref{fig:n_cutoff_heatmap_QEH}, some 'heat' is lost, signified by the lighter and more yellow colours, and some 'heat' is redistributed from the lower left quadrant of the heatmap to the upper right corner of the heatmap, thus illustrating the importance of material dependent screening. 
This heat restribution is similar to how heat was redistributed for the binding energies in going from Fig.~\ref{fig:E_b_analytic} to \ref{fig:E_b_ab_initio}. 
The effect of material dependent screening can also be seen in Fig.~\ref{fig:n_vs_m} where the strong correlation between the effective masses and cutoff densities observed in Fig.~\ref{fig:n_vs_m_analytic} is reduced substantially by the material dependent screening in the \emph{ab initio} results in Fig. \ref{fig:n_vs_m_QEH}. However, there is still a weak correlation between the effective mass and cutoff densities also in the \emph{ab initio} results, as will be discussed further below.

It is interesting to compare the heat map for exciton binding energies in Fig.~\ref{fig:E_b_ab_initio} with the corresponding heatmap for superfluid cutoff density in Fig.~\ref{fig:n_cutoff_heatmap_QEH}. 
The three materials with highest superfluid cutoff densities in \ref{fig:n_cutoff_heatmap_QEH} are found in the upper right 3 by 8 rectangle (where screening is generally smaller), but in this rectangle there is a large variation between its columns. The reason for this is that, contrary to the exciton binding energy which is mainly determined by the screening, the superfluid cutoff density depends both on the screening but is also heavily influenced by variations in the effective mass. This can also be observed in Fig.~\ref{fig:n_vs_m_QEH}. Thus the reason for the low cut-off densities for bilayers with p-doped $\rm ZrS_2$ or $\rm HfS_2$ is the small effective hole mass of these materials. 

\subsection{Identifying the top performers}
Finally, we now put our attention towards pointing out the most promising facilitators of exciton superfluidity. From our investigation of bilayers with direct band gaps the three bilayers, which are the best facilitators of exciton superfluidity are n-doped $\rm PbO_2$ combined with p-doped $\rm PtSe_2$, $\rm PdS_2$, or $\rm CrO_2$ with cut-off densities of $20\cdot 10^{12}\,\mathrm{cm^{-2}}$, $19\cdot 10^{12}\,\mathrm{cm^{-2}}$, and $18\cdot 10^{12}\,\mathrm{cm^{-2}}$, respectively. This aligns with the suggestion from the $E_b$-results that \acp{TMO} are promising platforms for exciton condensation. Out of bilayers exclusively with \acp{TMD} the chromium based compounds with heterostructures combining CrS$_2$, CrSe$_2$ and CrTe$_2$ have generally the highest cutoff densities with n-CrS$_2$,p-CrS$_2$ and n-CrSe$_2$,p-CrS$_2$ being the top performers at a cut-off density of $15\cdot 10^{12}\,\rm cm^{-2}$. It should be noted that many of the identified top performers, especially the TMOs, can be challenging to grow in monolayer form. However, single to few monolayer samples of the chromium dichalcogenides \cite{sun2021, zhang2021, li2021, habib2019, Wang2018} as well as  $\rm PtSe_2$ \cite{wang2015} have been reported in literature. Furthermore, many of the Cr compounds are reported to exhibit ferro- or antiferromagnetic ordering in the $T$ and $T'$ structures. We have therefore focused on the $2H$ phase in this work which is less prone to magnetic ordering \cite{hasan2023}.
\begin{figure*}[ht!]
\centering
\begin{subfigure}[t]{0.93\textwidth}
    \caption{Cutoff densities obtained by solving the gap equation with a model input interaction calculated with fixed $\epsilon^i_M = 2$.}
    \includegraphics[width=\textwidth]{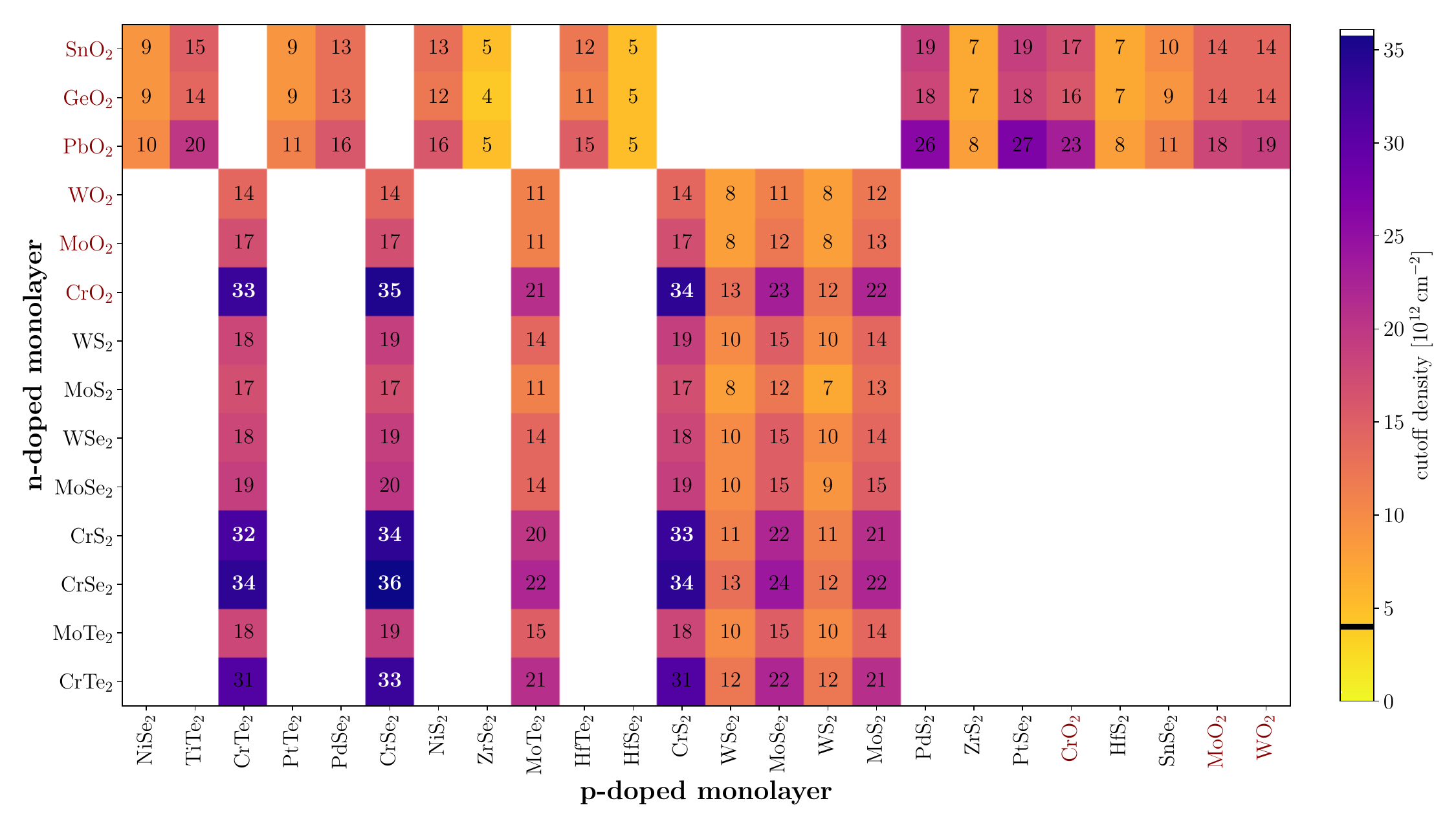}
    \label{fig:n_cutoff_heatmap_analytic}
\end{subfigure}
\hfill
\begin{subfigure}[t]{0.93\textwidth}
     \caption{Cutoff densities calculated obtained by solving the gap equation with \emph{ab initio} input parameters.}
    \includegraphics[width=\textwidth]{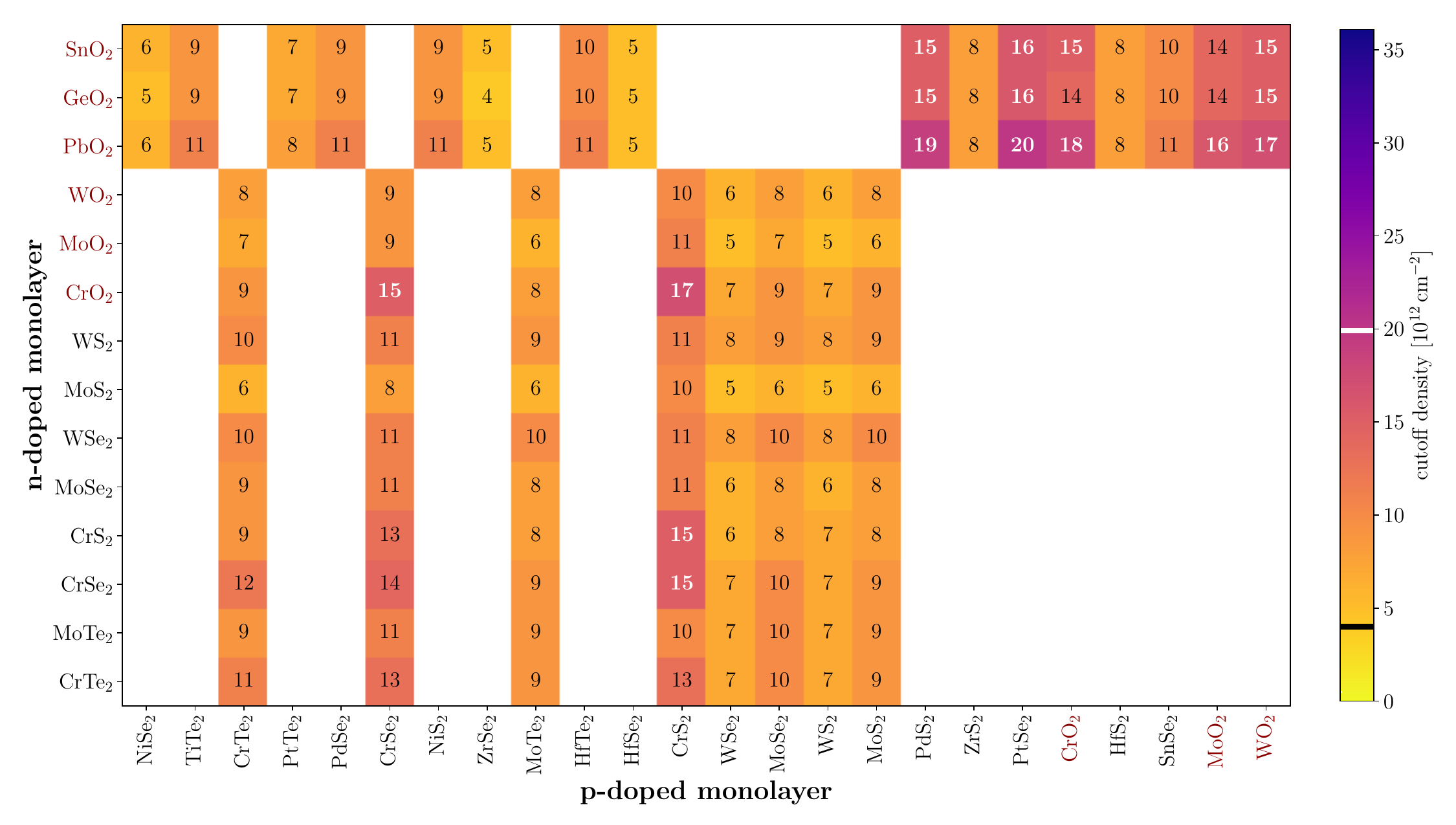}
    \label{fig:n_cutoff_heatmap_QEH}
\end{subfigure}
\caption{Heatmaps of cutoff densities (i.e. maximal carrier density where the gap disappears, see also Fig. \ref{fig:gap-analytic} and \ref{fig:gap-qeh}) as calculated by solving the gap equation for the bilayer systems using two different methods for calculating the effective model interaction. The sorting of materials is the same as in Fig. \ref{fig:E_b} and is determined from the values of the exciton binding energies from the \emph{ab initio} calculations (upper right corner bilayers had the highest $E_b$-values). \Acp{TMO} are written in dark red and \acp{TMD} are written in black. The top performer bilayers for each calculation method are highlighted with white labels. Since we required the interlayer bandgap to be direct for the gap equation calculation, 200 out of 336 bilayer combinations are blank in the heatmap.}
\label{fig:n_cutoff_heatmap}
\end{figure*}

\begin{figure*}[ht]
\centering
\begin{subfigure}[t]{0.495\textwidth}
    \caption{Cutoff densities calculated with fixed $\epsilon^i_M = 2$.}
    \includegraphics[width=\textwidth]{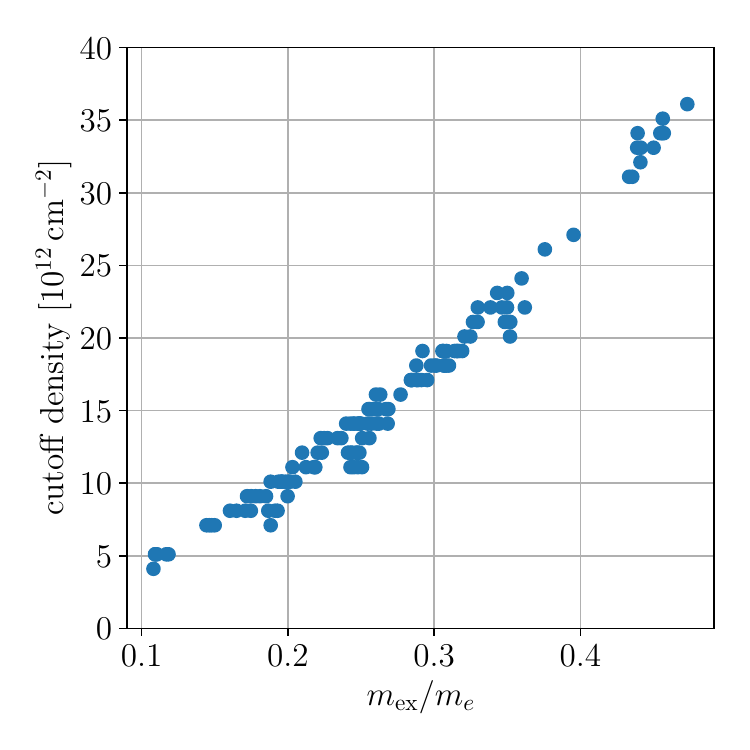}
    \label{fig:n_vs_m_analytic}
\end{subfigure}
\hfill
\begin{subfigure}[t]{0.495\textwidth}
    \caption{Cutoff densities calculated using the \emph{ab initio} input parameters}
    \includegraphics[width=\textwidth]{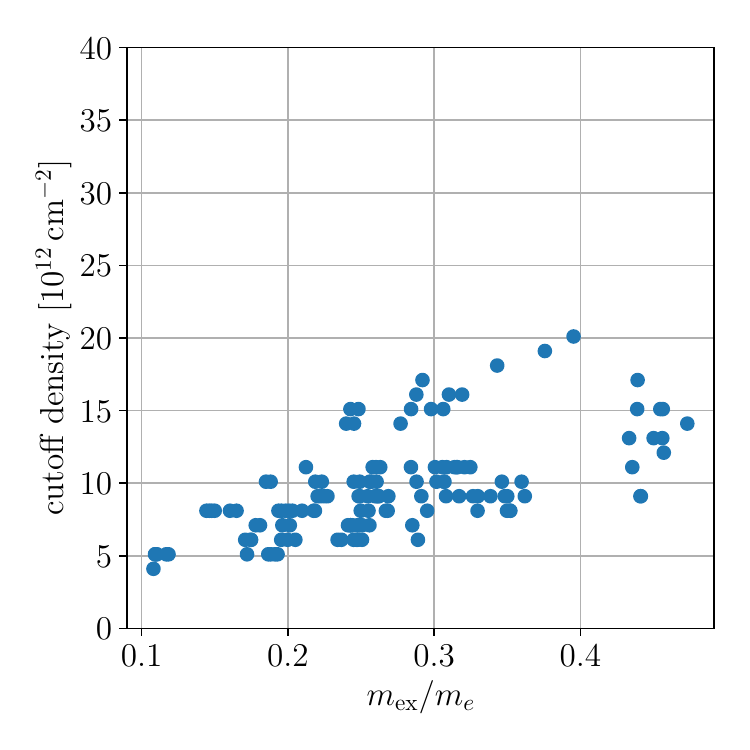}
    \label{fig:n_vs_m_QEH}
\end{subfigure}
\caption{Cutoff densities versus effective exciton masses with two different methods. See also caption to Fig. \ref{fig:n_cutoff_heatmap}.}
\label{fig:n_vs_m}
\end{figure*}
Due to the large influence of the effective mass on the superfluid cutoff density one could speculate that some material combinations in the lower right column of Fig. \ref{fig:n_cutoff_heatmap_QEH} that were excluded in our study, such as n-doped CrO$_2$ and  p-doped $\rm PtSe_2$, could have even higher cutoff densities. However, these materials have indirect bandgaps and therefore different valley degeneracy factors for the $n$ and $p$ doped layers which likely reduce their performance. Nevertheless, it would be interesting to extend the formalism to include also materials with indirect bandgaps.

\section{Conclusions \label{sec:conclusions}}
In this work we have calculated the exciton binding energies and superfluid properties of 336 \ac{TMO}/\ac{TMD} bilayer structures. First the equilibrium interlayer distance was evaluated using a $z$-scan approach. Then the exciton binding energies were computed using a Wannier Mott model and the superfluid properties (specifically maximal gap and condensate fraction) were investigated by solving a mean-field gap equation while accounting for screening using first principles. Because the screening is highly material dependent, we found that \emph{ab initio} treatments are needed to accurately predict magnitudes and trends in exciton binding energies and superfluid properties. Furthermore we outlined how the effective parameters for the superfluid gap equation, including the relevant screening channels, can be computed from first principles using the open source codes \ac{QEH} (Refs. \cite{Andersen2015, QEH_documentation}) and GPAW (Refs. \cite{GPAW_orig, GPAW_review}. Our study suggest that \acp{vdWH} with \acp{TMO}, explicitly n-doped $\rm PbO_2$ combined with p-doped $\rm PtSe_2$, $\rm PdS_2$, or $\rm CrO_2$ have the highest superfluid cutoff densities. However, we also found pure \ac{TMD} heterostructures combining  CrS$_2$, CrSe$_2$ and CrTe$_2$ to be superior facilitators of superfluidity compared to the \ac{TMD} bilayer structures which have previously been highlighted in experiments \cite{Wang2019, ma2021}. 


\section{Code availability}
All code for reproducing the results of this article can be found at the GitHub Repository in Ref. \cite{github_repo}. At the repository we have also included further computational details and documentation covering the workflow steps outlined in Figure \ref{fig:workflow} in the main text. The BiEx code package used to solve the gap equation is available from the authors upon reasonable request.
\\
\section{Acknowledgements}
This work was supported by the European Union’s Horizon 2020 research and innovation program under the Marie Skłodowska-Curie grant agreement No. 899987. (EuroTechPostdoc2).
We acknowledge funding from the European Research Council (ERC) under the European Union’s Horizon 2020 research and innovation program Grant No. 773122 (LIMA) and Grant No. 951786 (NOMAD CoE). K. S. T. is a Villum Investigator supported by VILLUM FONDEN (grant no. 37789)

\bibliography{references.bib}

\newpage

\newpage
\end{document}